\documentclass[sigconf]{acmart}
\AtBeginDocument{%
  \providecommand\BibTeX{{%
    \normalfont B\kern-0.5em{\scshape i\kern-0.25em b}\kern-0.8em\TeX}}}

\setcopyright{rightsretained}
\copyrightyear{2022}
\acmYear{2022}
\acmDOI{10.1145/3514094.3534185}

\acmConference[AIES'22]{Proceedings of the 2022 AAAI/ACM Conference on AI, Ethics, and Society}{August 1--3, 2022}{Oxford, United Kingdom}
\acmBooktitle{Proceedings of the 2022 AAAI/ACM Conference on AI, Ethics, and Society (AIES'22), August 1--3, 2022, Oxford, United Kingdom}\acmDOI{10.1145/3514094.3534185}
\acmISBN{978-1-4503-9247-1/22/08}

\acmSubmissionID{aies182}

\begin{document}
\fancyhead{}

\title{From Explanation to Recommendation: Ethical Standards for Algorithmic Recourse}

\author{Emily Sullivan}
\authornote{Both authors contributed equally to this research.}
\email{e.e.sullivan@tue.nl}
\orcid{0000-0002-2073-5384}
\affiliation{%
  \institution{Eindhoven Artificial Intelligence Systems Institute}
  \institution{Eindhoven University of Technology}
  \streetaddress{P.O. Box 513}
  \city{Eindhoven}
  \country{The Netherlands}
  \postcode{5600 MB}
}

\author{Philippe Verreault-Julien}
\authornotemark[1]
\email{p.verreault-julien@tue.nl}
\orcid{0000-0001-5816-8947}
\affiliation{%
  \institution{Eindhoven University of Technology}
  \streetaddress{P.O. Box 513}
  \city{Eindhoven}
  \country{The Netherlands}
  \postcode{5600 MB}
}

\renewcommand{\shortauthors}{Sullivan and Verreault-Julien}

\begin{abstract}
People are increasingly subject to algorithmic decisions, and it is
generally agreed that end-users should be provided an explanation or
rationale for these decisions. There are different purposes that
explanations can have, such as increasing user trust in the system or
allowing users to contest the decision. One specific purpose that is
gaining more traction is \emph{algorithmic} \emph{recourse}. We first
propose that recourse should be viewed as a recommendation problem, not
an explanation problem. Then, we argue that the capability approach
provides plausible and fruitful ethical standards for recourse. We
illustrate by considering the case of diversity constraints on
algorithmic recourse. Finally, we discuss the significance and
implications of adopting the capability approach for algorithmic
recourse research.
\end{abstract}



\maketitle

\hypertarget{introduction}{%
\section{Introduction}\label{introduction}}

There is widespread agreement that providing explanations for model
decisions is important, especially for end-users. Such explanations can
help users gain trust in an otherwise opaque system. Explanations can
also spur user engagement on product-based platforms. However, there is
no one-size-fits-all box for successful explanations. Explanatory norms
differ depending on the stakeholder, the domain, and the specific goals
a user has \citep{Zednik:2021, Sullivan:2019b}. One specific explanatory
norm that is gaining more and more traction is \emph{algorithmic
recourse}
\citep[e.g.][]{Joshi:2019, Karimi:2020, Karimi:2021, Konig:2021, Lakkaraju:2021a, Rasouli:2021, Ustun:2019, Venkatasubramanian:2020}.

Algorithmic recourse was borne out of counterfactual explanation methods.
\citet{Wachter:2018} highlight three uses for counterfactual
explanation: i) answer why a certain decision was reached, ii) provide
the user with grounds to contest the decision, and iii) provide the user with actionable changes to reverse the decision. While
Wachter et al. argue that counterfactual explanation can satisfy all
three, recent work suggests otherwise \citep[e.g.][]{Russell:2019b}.
Models can make decisions based on immutable features, which may satisfy
(i) and (ii), while failing to satisfy (iii). Since algorithmic recourse
is concerned with the specific project of providing users with an
\emph{actionable} counterfactual explanation, immutable features prevent
users from getting feasible and actionable advice on what changes they
could implement to get a new decision.

There are clear benefits from the user's perspective for recourse and
some have argued for its ethical value \citep{Venkatasubramanian:2020}.
Recourse seems especially important in domains where algorithmic systems
are part of decision pipelines that greatly affect people's lives, such
as granting a loan, sentencing decisions in a judicial system context,
college admissions and more. Nevertheless, as \citet{Venkatasubramanian:2020} discuss, algorithmic recourse faces pitfalls. The important work
on fairly defining cost, distance, etc. is necessary. However, shared (ethical)
standards for constraining recourse counterfactuals in particular
directions are conspicuously absent, with papers approaching the problem
in different ways. Some focus on the desiderata of proximity
\citep[e.g.][]{Wachter:2018}, while others highlight the need for
sparsity \citep[e.g.][]{Grath:2018} or for user input for specific
feature constraints \citep[e.g.][]{Venkatasubramanian:2020}, and 
others emphasize the need for diversity \citep[e.g.][]{Mothilal:2020}.

While we do not provide an all things considered ethical argument that algorithmic recourse is the best way to approach the problems of opaque systems that make highly impactful decisions, we seek to make progress on how to best constrain algorithmic recourse---assuming recourse is desirable---by providing an ethical framework that helps design recourse recommendations. Accordingly, proposing ethical standards for recourse does not imply letting designers and suppliers of artificial intelligence systems off the hook. Algorithmic decisions do not become exempt of other ethical standards because of the presence of recourse. This work makes three contributions:

\begin{enumerate}
\item
  Recasting algorithmic recourse as a \emph{recommendation} problem, not
  an explanation problem. Taking recourse seriously as a recommendation
  problem allows us to utilize insights from research programs on
  recommendation systems, which are largely siloed from questions in
  explainable AI. Moreover, it separates two distinct desiderata for
  algorithmic recourse: methods of generating or extracting
  counterfactuals and how to \emph{explain} counterfactual information
  to users. Once we solve which recommendations are necessary for
  recourse, then we can ask the explanatory question about how to best
  explain these recommendations to users. It may turn out through user
  studies that providing recourse recommendations is more successful
  through a different explanatory framework besides counterfactuals.
\item
  Providing ethical standards (via the capability approach) that can guide research on how best to constrain algorithmic recourse toward feasibility and the well-being of users.
\item
 As a case study, we use the capability approach as grounding the
  value of diversity for recourse recommendations. We highlight gaps
 in current research and suggest paths forward by taking inspiration from the role of diversity in recommendation systems.
\end{enumerate}

We hope that this work contributes to establishing plausible and
fruitful ethical standards for recourse recommendations. 

Section 2
argues that recourse should be viewed as a recommendation problem, not an
explanation problem. In section 3 we introduce the capability approach
and make the case for its descriptive and normative adequacy. Section 4
looks at diversity constraints on recommendations to illustrate the
usefulness of the capability approach and viewing recourse as a recommendation problem. We discuss several topics of potential
significance for recourse research in section 5.

\hypertarget{algorithmic-recourse-from-explanation-to-recommendations}{%
\section{Algorithmic recourse: from explanation to
recommendations}\label{algorithmic-recourse-from-explanation-to-recommendations}}

\hypertarget{recourse-as-an-explanation-problem}{%
\subsection{Recourse as an explanation
problem}\label{recourse-as-an-explanation-problem}}

People are increasingly subject to algorithmic decisions, with an
increased use of `black-box' models. This presents a challenge and need
for explainability. Explainable AI can increase users' trust in the
system, aid developers in building more robust and reliable models, and
more. Moreover, regulations like the General Data Protection Regulation
(GDPR) and the Artificial Intelligence Act (AIA) discuss the importance
of end-users receiving an explanation or rationale for decisions
involved in algorithmic processing. This has spurred a flurry of
development of different methods and approaches to explaining black-box
models.

One explanatory approach that has gained significant traction is
counterfactual explanation (CE). CEs provide answers to
\emph{what-if-things-had-been-different} questions. The claim is that
understanding the modal space of a model can serve as a way to explain
and provide understanding of the model's decision boundary. One of the
benefits of CE is that building a proxy model, that is necessary for
other feature importance methods, need not be necessary
\citep{Mothilal:2020}. Instead, CEs probe the black-box model by
changing various inputs to see what changes would lead to a change in
the output.

As we have seen, \citet{Wachter:2018} highlight three uses for CEs.
Ethicists and those interested in algorithmic fairness have especially
latched onto (iii)---how CEs can provide users with actionable advice to
reverse the outcome---known now as \emph{algorithmic recourse}.
\citet[p. 10, emphasis in original]{Ustun:2019} define algorithmic recourse
``as the ability of a person to change the decision of the model through
\emph{actionable} input variables {[}\ldots{]}''.

Since recourse was borne out of CE, recourse itself has been understood
as a type of explanation method, especially salient in domains where
algorithmic systems are part of decision pipelines that greatly affect
people's lives. In these contexts, when users are given a negative or
unfavorable decision, advice on how to get a different result in the
future is top of someone's mind. Thus, a recourse explanation seems most
suitable.

While explanations can serve a number of different goals, like
transparency and trust
\citep{Lipton:2018, Sullivan:2019b, Tintarev:2007}, explanation first
and foremost has epistemic aims, like filling knowledge gaps and
enabling understanding \citep{Friedman:1974fk, Grimm:2010rt}. As such,
most works look at recourse through the lens of an explanation problem,
where the evaluative goals center around the epistemic goals of
explanation, such as understanding the model and its decision boundary
\citep{Wachter:2018}. For example, \citet{Ustun:2019} describe recourse
as a type of actionable CE. \citet{Mothilal:2020} evaluate their method
of generating recourse counterfactuals with other XAI methods,
specifically LIME \citep{Ribeiro:2016}, to show that recourse
explanations can provide users with understanding of the decision
boundary.

However, we propose that conceptualizing recourse as an explanation
problem is ill-suited. As we explain in the next section, the goals of
explanation are distinct from the goals of providing users with
actionable information. While in some cases the same counterfactual can
explain \emph{and} provide actionable information to reverse a decision,
it is not by virtue of the counterfactual's \emph{explainability} that it
provides actionable information. Instead, we propose that algorithmic
recourse is best understood as a \emph{recommendation} problem and that
doing so has the promise of improving metrics and methods for
algorithmic recourse.

\hypertarget{recourse-as-a-recommendation-problem}{%
\subsection{Recourse as a recommendation
problem}\label{recourse-as-a-recommendation-problem}}

CE methods generate counterfactuals by making small changes to input variables
that result in a different decision. Counterfactual generation serves as
an explanation method because finding the smallest changes that would
flip a decision tells us important information regarding how a model
made its decision \citep{Stepin:2021}. However, sometimes
counterfactuals involve changing features that are immutable, or mutable
but non-actionable \citep{Karimi:2021}. Immutable features are those
that \emph{cannot} change, for instance someone's race. Mutable features
can change, but not because of a direct intervention on them. Someone's
credit score may change as a result of debt repayments, but it is not
possible for someone to intervene on her credit score. For this and
other reasons, the goals of explanation \emph{simpliciter} can come
apart from the goals of actionable information important for algorithmic
recourse. In this section, we discuss that explanation is possible
without recourse and that recourse is possible without explanation,
indicating that recourse is better understood as a recommendation
problem.

\hypertarget{explanation-without-recourse}{%
\subsubsection{Explanation without
recourse}\label{explanation-without-recourse}}

The first reason why algorithmic recourse is ill-suited to be an
explanation problem is that CE is possible without recourse
\citep{Ustun:2019, Venkatasubramanian:2020}. Consider the difference
between the following counterfactual explanations for a loan decision
discussed above: ``If you had less debt, then the loan would have been
approved,'' versus ``if you were younger, then the loan would
have been approved.'' The former CE gives the end-user recourse, while
the latter does not. It is not actionable advice for someone to become
younger, though it is actionable advice for someone to pay off some of
their debt. Moreover, in criminal justice cases, using a simplified
model based on COMPAS data \citep{Angwin:2016, Dressel:2018}, CE methods
found that race is often one of the more common features that would
reverse a risk categorization \citep{Mothilal:2020}. But again, since
race is immutable, it cannot be a recourse explanation but \textit{is} an explanation of the model's decision. Along these lines, \citet{Karimi:2021a} make a distinction between contrastive
explanations and consequential recommendations, the latter being a
subset of the former. The idea is that recommendation requires
information on the causal relationship between inputs, while explanation
just requires information regarding the relationship between the model and its inputs. If recourse requires a consequential
recommendation---which \citet{Karimi:2021a} argue is the case---then
again explanation is possible without recourse, especially since the
causal relationship between inputs involves a heavier burden to satisfy
(more on causation in section 5).

\hypertarget{recourse-without-explanation}{%
\subsubsection{Recourse without
explanation}\label{recourse-without-explanation}}

Even though most works discuss that a CE need not entail recourse, recourse can still be first and foremost an explanation problem. Recourse could be understood as
a specific \emph{type of explanation} that is actionable \citep{Karimi:2021a, Ustun:2019}. However, a less appreciated
distinction is that it is possible to have a recourse counterfactual
that fails to be an \emph{explanation}.

\citet{Barocas:2020a} highlight a notable difference between
principle-reasons explanations and recourse explanations. The former provide the data-subject with information regarding which features
serve as a justification or rationale against the decision, while recourse
explanations provide helpful advice \textit{without} the decision subject
learning about the features that were ``crucial marks against'' them.
Recourse serves a practical purpose of giving decision subjects guidance
for the future. Thus, having the most salient explanation that can
answer \emph{why} a model made its decision---or the rational for the
decision---can come apart from providing users with
\emph{recommendations} on how to reverse the decision. Consider again
the example of a recidivism classifier or loan decision algorithm as
discussed above. It very well might be that the immutable factors were
the more discerning factor for the decision. In this case, a recourse
`explanation' focusing on actionable factors becomes epistemically
misleading since the most discerning reason for the model's decision is
hidden. The user does not have access to the central difference-makers
of the model's decision, and thus would fail to really understand the
model. 

Conceptualizing recourse as a type of explanation can also mask bias. Explanation methods are used for auditing the fairness of models \citep{Lipton:2018}, with one central source of bias resulting from models using immutable features in a problematic way. Since recourse disregards counterfactuals that involve immutable features, recourse has the potential to mask bias and be epistemically misleading.

\hypertarget{recourse-as-recommendation}{%
\subsubsection{Recourse as
recommendation}\label{recourse-as-recommendation}}

The chief goals of model explanation center around providing users with
understanding the rationale of the model's decisions. Recommendation
systems, on the other hand, have a different primary goal.
They seek to help users with selecting a subset of
items that are among an ever-growing list of possible items by creating user profiles that are
continuously updated to aid in filtering the most relevant items for
users. As such, recommendation systems explore a specific relationship between a user and the model that is not mirrored in more traditional explainability questions regarding why a black-box model made a decision. 

The difference between recommendations and explanations can be
subtle in some contexts. Often recommendation systems also provide
explanations to users as to \emph{why} they are seeing the
recommendations that they do. However, the recommendations and the
explanations of recommendations are distinct. Our proposal is that
algorithmic recourse stands to benefit from such a distinction. The
purpose of generating the list of actionable advice is distinct from
explaining this advice and explaining the model's decision boundary. 

The relationship between recourse and recommendations has not gone
unnoticed. There has been work that takes insights from algorithmic
recourse to improve recommendation systems \citep{Dean:2020}. And those
working on recourse make the explicit connection that recourse is
similar to recommendation systems \citep{Mothilal:2020}. However,
\citet{Mothilal:2020} stop short of casting the goals of recourse to be
recommendation goals, since they evaluate their recourse model as if it
was an explanation problem, as discussed above. \citet{Karimi:2021a}
distinguish between two types of questions for recourse. (Q1)
explanatory questions, like ``why was I rejected for the loan?'', and
(Q2) recommendation questions, like ``What can I do to get the loan in
the future?'', where answers to Q2 questions provide ``consequential
recommendations.'' However, this terminology aims to point out a
difference in causal presuppositions needed for counterfactual
generation. They do not explicitly reconceptualize recourse as dealing
with the class of problems found in the recommendation systems
literature.

Our contribution is to explicitly conceptualize recourse as a
\emph{recommendation} problem akin to those problems facing
recommendation systems and not as an explanation problem. The unique
feature of algorithmic recourse is not explanation, but rather giving
advice and finding a subset list of actions from a large possible subset
of actions (i.e.~recommending). It is our contention that shifting the
dialectic away from algorithmic recourse as an explanation problem to a
recommendation problem will improve recourse recommendations as well as
help to make sure that algorithmic recourse is not used in ethically or epistemically misleading ways. It shifts the focus away from explainability to a more user-modelling perspective regarding the interplay between user-preferences and capabilities and the model.

Once we solve which recommendations users should have such that recourse
is possible, then we can ask the question how best to explain or convey
this information to users. This may be through counterfactuals, or it
may turn out through user studies that providing recourse
recommendations is more successful through a different explanatory
framework. An added benefit of considering recourse as a recommendation
problem is that it allows us to utilize insights from a rich research
program in recommendation systems that is still largely siloed from
questions in XAI. Moreover, conceptualizing recourse as a recommendation
problem allows us to utilize particular ethical tools---like the
capability approach---to guide research in filtering counterfactuals
that respond well to users' capabilities even if they are far removed
from the model's decision boundary.

\hypertarget{ethical-standards-for-recommendations-the-capability-approach}{%
\section{Ethical standards for recommendations: the capability
approach}\label{ethical-standards-for-recommendations-the-capability-approach}}

\hypertarget{the-ethical-standards-of-recommendations}{%
\subsection{The ethical standards of
recommendations}\label{the-ethical-standards-of-recommendations}}

In theory, recourse has ethical appeal through purportedly promoting agency and autonomy.
\citet{Venkatasubramanian:2020} provide some general ethical standards
for algorithmic recourse by arguing that it is a modally robust good
\citep[see][]{Pettit:2015}. Robust goods deliver benefits in a range of
actual and counterfactual circumstances. For example, the robust good of
honesty provides the benefit of truth-telling not only on one specific
occasion, but on many occasions. According to this view, we value robust
goods because they deliver benefits in various circumstances.

Venkatasubramanian and Alfano hold that someone who has recourse enjoys
a capacity to obtain decisions across a range of circumstances and not
in a coincidental or piece-meal fashion. That person can reasonably
expect that she will be able to obtain a decision and will not be
subject to other people's discretionary power or to changing situations. This
is crucial for exercising what Venkatasubramanian and Alfano call
`temporally-extended agency', namely the capacity to pursue long-term
plans. This sort of agency is important because algorithmic decisions
are often a means among a chain. A person seeking a loan to buy a car,
they say, may do so in order to take a well-paying job which itself is a
means to care for her family. The implications of being denied a loan
are thus more far-reaching than simply not being able to obtain the
immediate goods or services the loan is for.

While Venkatasubramanian and Alfano provide both consequential (Pettit's
framework) and deontological (based on human dignity) reasons to value
recourse, how these foundations relate to specific constraints on
recommendations and how they may help comparing them remains
unclear. They discuss a variety of issues, for instance changes to
classifiers over time, and importantly convey that these issues need to
be resolved for algorithmic recourse to live up to its ethical promise.
Other works on recourse have differed in their approach to the
evaluation of constraints, picking and choosing which are necessary or interesting for their specific study, with some of the above
concerns in mind.\footnote{For a survey, see \citet{Karimi:2021a}.} However, no principled ethical framework is currently guiding the design of recourse recommendations. 
In order to make progress on algorithmic recourse, we need to make
progress on delineating which reasons may justify adopting some
constraints over others. We need ethical standards that can do this
work. We propose that that the capability approach provides such plausible
and fruitful standards. First, we introduce the capability approach
and then illustrate its relevance by considering one particular
constraint: diversity (section 4). In section 5, we discuss the more general
significance of the capability approach for recourse research.

\hypertarget{the-capability-approach}{%
\subsection{The capability approach}\label{the-capability-approach}}

The capability approach, initially developed by Amartya Sen
\citetext{\citealp{Sen:1979, Sen:1980fk, Sen:1992}; \citealp[see
also][]{Nussbaum:2000, Robeyns:2017a}}, is a normative
framework which characterizes the normative space of evaluation in terms
of \emph{functionings} and \emph{capabilities}. According to the
capability approach, we should make
interpersonal comparisons or assess states of affairs on the basis of these two core concepts.
Functionings are `beings'---ways of being, like being healthy or
educated---and `doings'---activities, like coding or cycling ---people may be or undertake. Having an appropriate set of functionings is
``constitutive of human life'' \citep[p. 39]{Robeyns:2017a}; what makes up
and gives value to human life are the `beings' and `doings' people
achieve. Capabilities are the real freedoms, or opportunities, people
have to achieve functionings. Here, `real' underlines that having a
capability goes beyond having a merely formal possibility. It requires
having the resources (broadly construed, e.g. income, credentials,
social network, etc.) to effectively achieve chosen functionings.
Another important claim of the capability approach is that the
capabilities people have depend on \emph{conversion factors}, namely the
differential capacity to convert resources into functionings. With equal
resources, different people will not always have the same capabilities.
Other things being equal, a person who suffers from depression will need
more resources to achieve the same level of motivation as someone
without depression. Conversion factors can be personal (e.g.~a
disability), social (e.g.~being discriminated), or environmental (e.g.~
the climate) and can be intertwined. Acknowledging conversion factors is
important for ethical evaluation because it urges caution in equating
resources with well-being.

\begin{figure}[h]
    \centering
    \includegraphics[width=\linewidth]{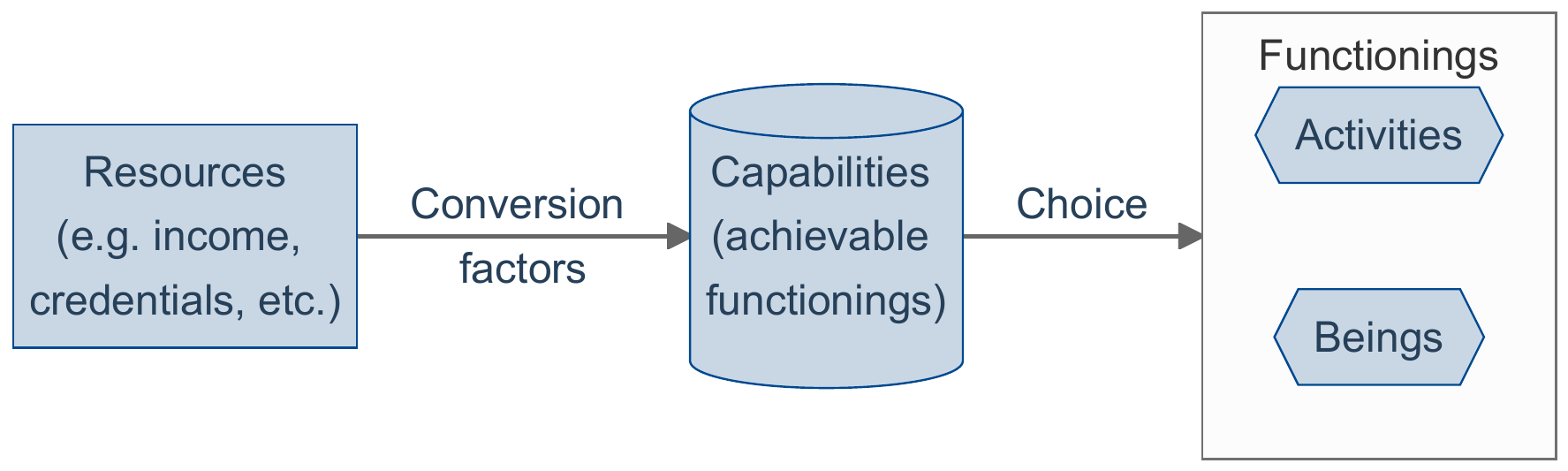}
    \caption{Schematic overview of the capability approach. Resources are converted into capabilities and people choose which functionings to realize from their set of capabilities.}
    \Description{A schema of core concepts of the capability approach.}
  \end{figure}

The notion of capability aims to distinguish between what is
\emph{actually realized} (functionings) versus what \emph{could effectively} be
realized (capabilities) if people wanted to. As figure 1 illustrates, resources are
converted into capabilities, effectively possible but unrealized
functionings. From that capability set, a
person then chooses which functionings to actually achieve. For instance, someone
may have the capability to cycle, yet never do it. That person may
opt for moving about using public transportation. Again, what matters is
the real freedom people have to achieve a combination of functionings.

\begin{figure}[h]
    \centering
    \includegraphics[width=\linewidth]{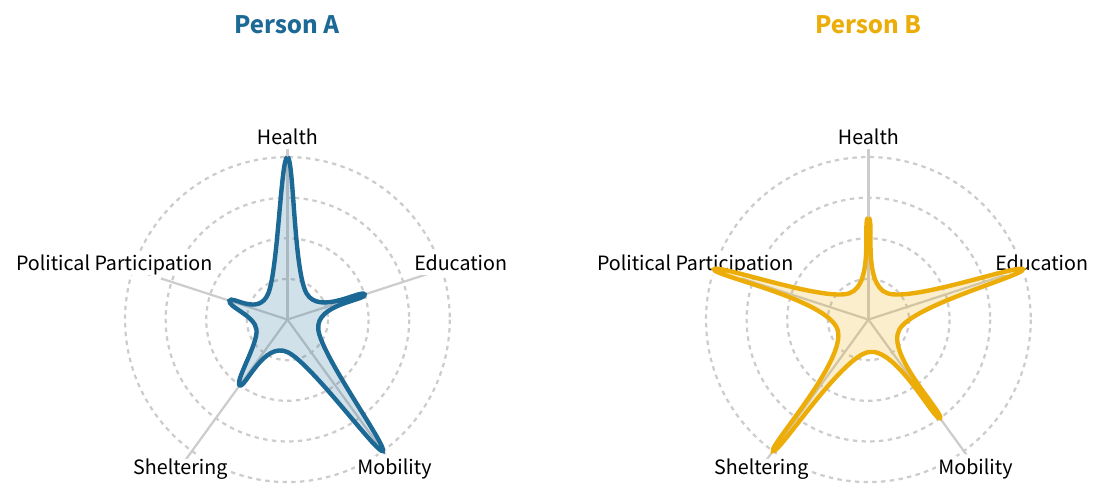}
    \caption{Interpersonal comparison of capabilities. The capability
    approach holds that we should compare people's advantage in terms of the
    capabilities and functionings they have. Figure made with \href{https://flourish.studio}{Flourish}.}
    \Description{A radar chart comparing the capabilities of two persons.}
  \end{figure}

A capability set is the set of alternative functionings people can
achieve. For instance, let us consider the capabilities to be healthy,
educated, mobile, sheltered, and participate in politics (see
figure 2). Different people may have different capability sets, due e.g.
to conversion factors, and thus have a differential real freedom to
achieve the related functionings. For instance, Person A might have a
greater capability for health than B, but B might be advantaged in terms
of education, perhaps because of the social environment. The capability
approach holds that interpersonal comparisons should be made in terms of
capabilities and functionings.

While figure 2 represents a `static' capability set, in reality there
are often trade-offs between capabilities. As figure 3 shows, having
more of one capability may sometimes have positive or negative effects on
the capability set. Using more resources in order to gain an increased
capability in terms of education might have a negative effect on the
capability for health, which in turn might reduce one's mobility. More
education, however, might contribute positively to political
participation. People
face similar trade-offs all the time when considering the real
opportunities they have. Some could become a scientist or a rock
musician, but achieving both is not always effectively possible.\footnote{One notable exception is Brian May, guitarist of the famous band Queen, who received a PhD in astrophysics in 2007.}

\begin{figure}[h]
    \centering
    \includegraphics[width=\linewidth]{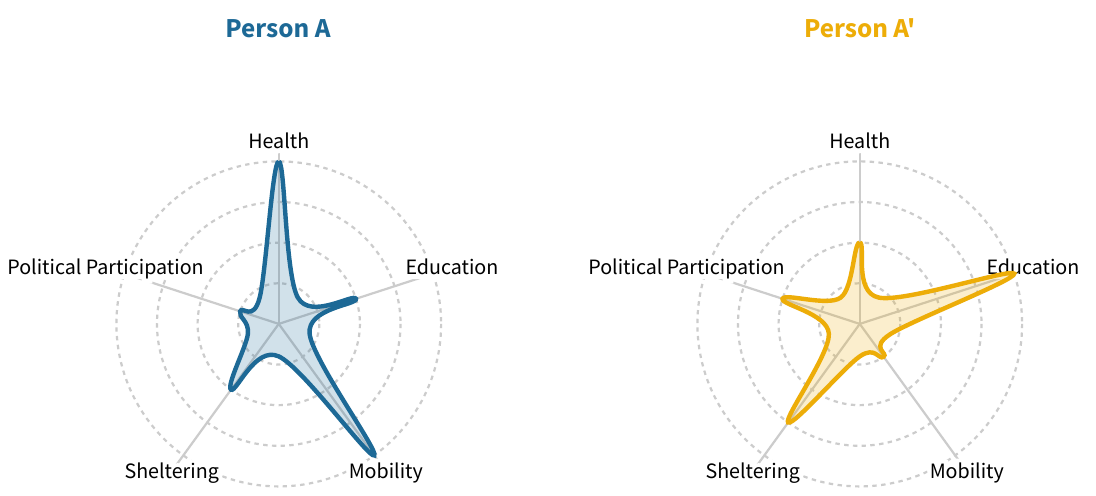}
    \caption{Trade-offs between capabilities. Resource allocation for one capability can have an influence on other capabilities. Figure made with \href{https://flourish.studio}{Flourish}.}
    \Description{A radar chart illustrating possible trade-offs between capabilities.}
  \end{figure}

Capabilities help capture the idea that the freedom to achieve
certain beings and doings is of utmost moral value. A person's
well-being is constituted by what is ultimately good for that person. As
\citet[231]{Sen:2009fk} notes, any ethical or political theory must
select an `informational basis', viz.~features of the world that help to
assess well-being and injustice. The capability approach contrasts with
alternative theoretical frameworks by submitting that these features are
the capabilities people have reason to value instead of, for instance,
pleasure or resources. This broadens the informational basis insofar as
information about resources or rights can be legitimately used to
compare well-being. How to determine the relevant capabilities for the
purpose of normative assessment is context-dependent. It can be used for
assessing individual well-being, evaluating social states of affairs, or
policy-making \citep{Robeyns:2021}. It is an influential framework that
has been used in fields such as human development \citep{Gaertner:2006},
poverty \citep{Alkire:2015}, mental health \citep{Simon:2013},
technology \citep{Oosterlaken:2009, Zheng:2009}, or education
\citep{Walker:2007}. One famous use of the capability approach is within
the United Nations Development Programme's \emph{Human Development
Reports}, in particular the Human Development Index.\footnote{http://hdr.undp.org/en/content/human-development-index-hdi}
For the purpose of assessing and comparing human development between
countries, using indicators such as life expectancy or the level of
education may target adequate capabilities. But for assessing whether
older people have mobility through public transport, looking at
residential density and physical functional capacity would be more
relevant \citep{Ryan:2015}.

\hypertarget{recommendations-and-the-capability-approach}{%
\subsection{Recommendations and the capability
approach}\label{recommendations-and-the-capability-approach}}

The capability approach provides plausible and fruitful ethical
standards for recourse recommendations because it is descriptively and normatively adequate.

\hypertarget{descriptive-adequacy}{%
\subsubsection{Descriptive adequacy}\label{descriptive-adequacy}}

The capability approach is descriptively adequate because it captures
the relevant features of recourse recommendations. Current formulations
of recourse have natural analogues within the capability approach.
Recourse can readily be understood as a functioning; it is the
\emph{activity} of obtaining a decision from a model. When someone
obtains a decision, that person achieves the functioning of recourse.
But recourse is also viewed as an `ability' or as something that a
person has the `capacity' to do irrespective of whether they actually
achieve it or not. As such, recourse is also a capability; it amounts to
the real freedom to obtain a decision from a model. When someone has
recourse, that person \emph{would be able} to obtain a decision would
she choose to do so. Viewing recourse as a capability also explains the
widespread emphasis on actionability. Recommendations are those that
users could in principle, but not necessarily, achieve.

Although the notion of capability captures usage of recourse in the
computer science literature, it also stresses one underrated feature of
recourse, namely its connection to \emph{freedom}. Capabilities are a
type of freedom, in particular \emph{option-freedom}
\citep[see][pp. 102ff.]{Robeyns:2017a}. Options are what an agent can
achieve or realize. The freedom of options depends on two aspects: 1)
the agent's access to the options and 2) the options themselves. Some
people may face more obstacles (e.g.~different conversion factors) than
others to realize certain options, resulting in different access to options (1).
Option-freedom also depends on the number or quality of options
available (2). A person with more options has more
option-freedom than a person with fewer options.

For the purpose of recourse, recommendations (should) aim to give option-freedom. In fact, viewing recommendations as seeking to
promote option-freedom helps understand the aims of different
recourse methods. Some emphasize the importance of causal
possibility \citep[e.g.][]{Karimi:2021} and thus that people should have
the proper access to options (see sec 5.1 below for a critique). Others
draw attention to the options themselves by generating a large quantity
of options users can choose from \citep[e.g.][]{Mothilal:2020}. Adopting
the capability approach thus provides a rich description of what
recourse is, explain its usage and the
motivations behind specific recourse methods.

\hypertarget{normative-adequacy}{%
\subsubsection{Normative adequacy}\label{normative-adequacy}}

The capability approach is normatively adequate because it picks out 
relevant normative features for designing and assessing recommendations. 


First, it picks out an important moral feature of recourse recommendations, viz.~that people who can obtain a decision from an algorithm are in a better position than those
who are not. Recommendations that provide recourse
qua capability give them the real freedom to obtain decisions.
Insofar as we accept that one key metric of well-being is people's capabilities, it follows that promoting the capability of recourse will
also promote people's well-being. Second, the capability
approach provides a substantive, but flexible, evaluative framework to
design and compare recommendations. In particular, it provides the key
metric recommendations should optimize for, namely capabilities.
Consequently, good recommendations will be ones that fall within a
person's capability set. If a person does not have the capability to
achieve the recommendation, then that recommendation is not actionable
and, crucially, that person does not have recourse. When assessing
recommendations, we should thus pay special attention to whether people
have the capability to achieve them.

As we noted earlier, there are various reasons why we would consider recourse to be valuable,
for example because of its role in agency and autonomy. We do not deny
that those may ground the value of recommendations. In fact, our goal
is more modest: assuming we want recourse, what are fruitful ethical
standards for designing and assessing recommendations? One key advantage of the capability approach over alternative evaluative frameworks is that it broadens the informational basis. For instance, it takes into account people's preferences, but also incorporates information about their conversion factors and the (real) freedom people have to achieve functionings. As a result, recommendations that aim to promote capabilities can come apart from recommendations that aim to solely satisfy preferences.

To illustrate, suppose someone would like to receive a recommendation for obtaining a loan. Recommendations that aim to promote the satisfaction of preferences face several challenges. One of them is that it is not always possible to act on one's preferences. Someone born in Canada might have a preference for becoming President of the United States, but it is impossible to satisfy that preference. Only natural-born-citizens may become President. Likewise, giving users recommendations that they prefer, but are not actionable to them, will not contribute to their well-being. Another challenge is that since the recommendation process may itself contribute to shaping preferences, then the users' preferences become a moving target. We could assume that a user seeking a recommendation for a loan would prefer to obtain it and that, accordingly, the recommendation should help the person satisfy that preference. However, a recommendation may show that obtaining the loan could only be done through a difficult process. Even though the person would have the capability of achieving the recommendation, she might choose, or prefer, to not do so. The capability approach emphasizes that giving users the \emph{freedom} to realize a preference, not its actual satisfaction, is what matters for recourse. 

A last challenge is that since the preference-satisfaction framework is fundamentally individualistic, it fails to take into account structural constraints from the social environment. On the contrary, the capability approach can incorporate larger social complexities via conversion factors and by broadening the informational basis \citep[see, e.g., secs. 2.7.5 and 4.10]{Robeyns:2017a}. This then allows to take into account differences between groups (see section 5 below).

One specific constraint that falls out of the capability approach is
that recourse explanations should be diverse. In other words, in order
for users to increase their capabilities requires that they are given
more than one recommendation, and that these recommendations are in an
important sense distinct. In the next section, we look closely at the
constraint of diversity and the value it has for algorithmic recourse.

\hypertarget{the-value-of-diverse-recommendations}{%
\section{The value of diverse
recommendations}\label{the-value-of-diverse-recommendations}}

In what follows, we show the fruitfulness of conceptualizing algorithmic
recourse as a recommendation problem and the fruitfulness of the
capability approach by taking a close look at the constraint of
\emph{diversity} on algorithmic recourse. \citet{Wachter:2018} discuss
the importance of providing diverse recourse recommendations, with many others
agreeing \citep{Mothilal:2020, Russell:2019b}. However,
detail about \emph{why} diversity matters and how diversity constraints
specifically can overcome some of the problems facing recourse is
lacking. Moreover, diversity constraints are largely undervalued in
current research on algorithmic recourse. Only 16 of the 60 recourse
algorithms found in a recent survey \citet{Karimi:2021a} include
diversity as a constraint. And of the works that include diversity, several lack sufficient detail motivating their choice of
diversity metric. Meanwhile, the value of diversity in recommendation
systems is well documented with several research lines investigating the
best suited diversity metrics for specific use cases
\citep{Kunaver:2017}, as well as user perceptions and reactions to
diversity \citep{Castagnos:2013, Hu:2011, Tintarev:2013}.
\citet{Vrijenhoek:2021}, in their work on diverse news recommendation,
develop diversity metrics that reflect normative democratic values. In a
similar vein, the capability approach can serve as a motivation for
specific diversity metrics for algorithmic recourse.

\hypertarget{diversity-for-recourse-recommendations}{%
\subsection{Diversity for recourse
recommendations}\label{diversity-for-recourse-recommendations}}

Providing users with a diverse set of recourse recommendations is
currently motivated because of prevailing uncertainty in user
preferences \citep{Karimi:2020}. This problem has analogs to the cold
start problem in recommendation systems, where recommendations are
provided even when the system has little data regarding the user or
their behavior \citep{Schein:2002}. Providing users with a diverse
set of recommendations is one way to overcome the cold start problem
\citep{Kunaver:2017}. However, there are additional reasons for valuing
diversity besides uncertainty in user preferences. For example, in news
recommendation diversity can help with combating filter-bubbles
\citep{Loecherbach:2020}. Importantly, depending on the overall purpose of diversity, different diversity metrics are more or less suitable \citep{Vrijenhoek:2021, Sullivan:2019b}. Thus, the fact that
diversity in algorithmic recourse only seeks to address
uncertainty in user preferences narrowly constrains the choice of
diversity metrics. If diversity in recourse recommendations is valuable for other purposes---e.g. broadening one's capability set---then the choice of suitable diversity metrics will be notably different. 

The majority of works in algorithmic recourse understands diversity as a
type of distance or similarity metric between counterfactuals
\citep{Karimi:2021a}. While this approach may very well yield diverse
counterfactuals that help to overcome uncertainty in user preferences,
there are drawbacks. First, the similarity or distance function is
operative both in generating the list of possible counterfactuals and
also in selecting the diverse set. This can retain biases that result in
determining distance or similarity in the first place. However, the
value of diversity metrics is that they have the potential to counteract
this bias by considering other trade-offs. For example,
\citet{Dandl:2020} discuss diversity in relation to trade-offs between
different objectives, such as the number of feature changes, closeness
to the nearest observed data points, and plausibility according to a
probability distribution. They argue that exploring trade-offs improves
understandability and the number of options for the user compared to
other approaches that build in \emph{a priori} a weighted sum.
\citet{Mothilal:2020} also describe different trade-offs. They identify
proximity diversity and sparsity diversity. The former concerns the
distance and the latter the number of features that need to be changed
to reverse the decision.

Moreover, most current works on algorithmic recourse diversify recommendations post-hoc (i.e.
after initial counterfactual generation). However, as learned from work
in recommendation systems, post-hoc diversity methods face a problem
that if the initial generated list is not diverse, the diversity metrics
do little to help \citep{Kunaver:2017}. Making progress on the
effectiveness of diversifying recourse recommendations starts with
conceptualizing recourse as a recommendation problem and then learning
from the various methods of diversity discussed in recommendation
systems.

\hypertarget{capability-approach-and-diverse-recourse-recommendations}{%
\subsection{Capability approach and diverse recourse
recommendations}\label{capability-approach-and-diverse-recourse-recommendations}}

The capability approach not only tells us why diverse recourse
recommendations are valuable---because they increase the likelihood that
a user actually has the capability to have recourse---it provides a way of thinking about
ethical standards for diversity metrics. First, recommendations
are usually evaluated based on how accurate recommendations are for
fulfilling user preferences. However, the capability approach tells us that it is not preferences that should make up the evaluative space, but a user's
capabilities. This would entail that evaluating whether a recourse
recommendation is successful should not be geared toward
preference-satisfaction, but promoting capabilities. Second, following
the method of \citet{Vrijenhoek:2021}, we identify two key normative
themes that motivate how to diversify recourse recommendations. While it
is possible that the capability approach could motivate more
considerations of diversity, we highlight two that are currently missing from recourse diversity metrics.

\hypertarget{temporality}{%
\subsubsection{Temporality}\label{temporality}}

The capability approach highlights that capabilities have the potential
to be realized involving various trade-offs and time frames, with
\citet{Venkatasubramanian:2020} discussing the value of recourse as a
type of temporally extended agency. Recourse recommendations can account
for this temporal dimension by diversifying the time frame for realizing
a capability. For example, getting an additional educational degree may
take more time compared to other activities. Another aspect of
temporality is the time it might take before particular
capabilities become possible. For example, someone may have several capabilities that are only realizable
after their children become a certain age.

The diversity metric of temporality diversifies recourse recommendations
based on differences in user capability time frames. Current recourse
techniques account for aspects of temporality through a brute cost
function, with cost generally understood as a probability distribution for a given feature compared to others.
Diversifying over temporality focuses on another kind of cost: time.
It gives the user the ability to see for themselves the options for a
shorter versus longer time frame potentials. 

\hypertarget{resource-conversion}{%
\subsubsection{Resource conversion}\label{resource-conversion}}

The capability approach highlights that different people have different
conversion factors (i.e. the differential capacity to convert resources
into functionings). Equality of resources does not imply equality of
capabilities. Resource conversion diversifies
over a range of more or less resource intensive actions. While resource conversion shares many similarities with current cost metrics, the capability approach urges us to understand cost differently from the probability distribution method that is currently popular among
recourse algorithms. The probability method of cost assumes that
everyone has the same conversion factors. However, this is not the case. The capability approach
motivates diversifying cost to reflect the differences in users'
conversion factors. Gaining knowledge about a user's specific
conversion factors could improve the accuracy of recommendations, but diversifying on resource conversion is still valuable according to the capability approach to facilitate option-freedom.

\hypertarget{limits-of-diversity}{%
\subsubsection{Limits of diversity}\label{limits-of-diversity}}

Maximizing diversity and including a never-ending list of diverse
recommendations will not be successful for providing users with
actionable choices. There are a variety of trade-offs that we need to
consider when devising specific recourse recommendations. For example,
people can face `option overload' when there are too many live options
to choose from. As a result, adding yet another diverse recommendation
may actually reduce one's capability set since it makes it harder to convert a recommendation into an achievable functioning. Thus, it is important to engage in
user-study research concerning the number of recommendations that is
optimal. The length of the list could differ between users, with some
users achieving their goals with two options, while for others, five options may be optimal. The capability approach may help
in navigating how to handle such trade-offs. Specifically, user-studies
should be designed that seek to validate the extent to which one's capability
set is captured, instead of the feeling of trust the user has in the system. Additional options include getting user input regarding which diversity metrics they are interested in seeing for recourse recommendations. 

\hypertarget{significance-for-recourse-research}{%
\section{Significance for recourse
research}\label{significance-for-recourse-research}}

The capability approach provides a conceptual and normative framework
against which we can assess and compare different constraints and
proposals for recommendations. Naturally, it does not (and will not) settle
all disputes, but no theoretical framework can do that. But it is
important to at least agree on \emph{what terms} disputes should be
settled. These terms are that recommendations should promote people's
capabilities. As a result, we believe that the capability approach may
help define adequate optimization procedures besides diversity. In this section, we
present several implications that adopting the capability approach has
on current themes in recourse research.

\hypertarget{causality}{%
\subsection{Causality}\label{causality}}

Some recent work \citep[e.g.][]{Karimi:2021} emphasize the importance of
building causal models to provide actionable recommendations. One
benefit of causal models is that they can assess which features are
immutable or non-actionable in the sense of not being causally possible.
Counterfactual explanations may not provide actionable recommendations
if there is no causal path between the features the user would have to
intervene on and the decision. For instance, it is not causally possible
to increase one's level of education while reducing or keeping one's age
constant. This why Karimi et al.~propose a method for generating
``recourse through minimal interventions''. Minimal interventions aim to
minimize the cost of implementing a set of actions that would change the
decision.

Although causal possibility is certainly an important dimension of
actionability, even if we assume away the problem of having perfect
causal knowledge \citep[see][]{Karimi:2020}, the capability approach
allows us to see that we arguably need to broaden the causal lens. Capabilities (or lack thereof) do not always neatly fall
within the `causal' category. Recall that capabilities are best
understood as option-freedoms and that they are a function of the character of the options themselves and their access. One's route to achieving recourse may be more difficult and less
accessible. One particularly pressing problem is that there might be a self-selection bias when people opt for some recommendations over others because of incorrect
beliefs about what they can possibly do or not. Or, perhaps even more
worrying, people might self-select because of normative beliefs about
what they \emph{should (not)} do. A woman might not consider a
recommendation as actionable because it involves increasing her level of
education, which would be frowned upon in her community. Other
recommendations might be so burdensome as not falling within one's
capability set, yet still being causally possible.

Another issue is whether conversion factors
(personal, social, or environmental) can always be represented in causal
terms. For instance, power relations and social norms may all affect one's ability to convert resources in capabilities. Moreover, it is contentious that social categories such as gender or race can be viewed as a cause \citep{Bright:2015, Glymour:2014, Hanna:2020, Kasirzadeh:2021, Marcellesi:2013, Weinberger:2021b}. But even if factors such as those could be represented as having a
positive or negative causal influence, our point is simply that accurate
causal models need to address problems of possible causal break-down and the complexities surrounding the way conversion factors can be causally efficacious.

\hypertarget{proxies}{%
\subsection{Proxies}\label{proxies}}

One way of understanding the role of constraints for recommendation
algorithms is that they are \emph{proxies} for actionability. Reducing
the distance between the factual and the counterfactual instance that crosses the decision boundary is one typical constraint. Other
common constraints include `plausibility'
(i.e.~likely to be actually instantiated) or `sparsity' (i.e.~recommending
changes to as few variables as possible). Distance, plausibility, or
sparsity are all proxies for actionability. Furthermore, as discussed above, since it is in practice
difficult to build complete and accurate causal models
\citep{Karimi:2020}, current causal models are also
a proxy for actionability. 
Although not directly
determining actionability, all the above constraints are often taken to constitute good
approximations for actionable recommendations. 

The capability approach provides a normative
framework for assessing which proxies might better optimize the relevant
notion of recommendation, viz.~recommendations that people have the real
freedom to achieve. For instance, the Human Development Index considers
that income per capita, education level, and life expectancy are good
indicators of human development along with the capabilities people have in
different countries. From this, we could infer that people with more
income, education, or life expectancy will have a greater capability to
implement recommendations. The likelihood of providing a truly
actionable recommendation for people who score high on these indicators
should be greater. This is just one example of how recourse qua
capability could be inferred, albeit imperfectly, from proxies.
Fortunately, there is a significant literature on measuring capabilities
in education, health, etc.
\citep{Al-Ajlani:2020, Alkire:2015, Gaertner:2006, Oosterlaken:2009, Simon:2013, VanOotegem:2012, Walker:2007, Zheng:2009}.\footnote{See,
  e.g., \citep{Alkire:2013, Robeyns:2003, vanderDeijl:2020a} for discussions of challenges to measuring
  and operationalizing the capability approach.} Designers of
recommendations systems could find from other fields relevant proxies for providing recourse for various applications and contexts.

One key advantage of using the capability approach is that it helps
answer ex ante and ex post questions about recommendations. The first
is: What are the best proxies of people's \emph{current} capabilities?
This is directly related to actionability insofar as we want to provide
recommendations that people have the real freedom to achieve. Following
the capability approach, the answer to that question is that the
recommendation should fall within one's capability set. Providing
diverse recommendations is one important means to achieve that goal. But
the second, often underrated, question is: What recommendations would
most \emph{improve} people's lives? The capability approach would
suggest that recommendations that improve more people's capabilities are
the better ones. Consider again the case of the proxies for human
development (income, education, health). On that basis, we might
conclude that recommendations that would privilege acting on income,
education, and health may have the greater impact on people's
capabilities. Ceteris paribus, people with more income, education, or
health are typically freer to achieve functionings. This would suggest
to favor recommendations that have the greater ex post impact.

\hypertarget{tough-recommendations}{%
\subsection{Tough recommendations}\label{tough-recommendations}}

Some recommendations may be actionable yet be `far-fetched' in the sense
of too difficult or burdensome to achieve.
\citet[sec.~4.6]{Venkatasubramanian:2020} argue that we should refrain
from giving such recommendations. Although we agree that such
recommendations may not be relevant in many cases, the capability
approach suggests caution before a priori deciding that a recommendation
is too difficult or burdensome. First, classifying a recommendation as
too costly implies that we have sufficient information about users'
current capabilities. In many cases, this assumption does not hold, which
is also why recommendations should be diverse. Second, this may
unduly interfere with people's capabilities. Nudging or not providing
recommendations may affect the access to options as
well as the options themselves. For instance, people may come to believe
that acting on a recommendation is too hard for them, which might not
really be the case. Or, excluding recommendations may restrict the
quantity and quality of options people believe they have access to. In
any case, we should be very wary of allowing recommendations systems to
limit the availability of recommendations.

\hypertarget{strategic-manipulation}{%
\subsection{Strategic manipulation}\label{strategic-manipulation}}

One concern of recourse research is that users may try to strategically
manipulate algorithms. From the perspective of the capability approach,
it is unclear why `gaming the system' is a problem for users. If we want
to promote people's capabilities, giving people recommendations that
they may use for achieving functionings that they value would indeed
promote their capabilities. This may seem like a bug, but it is a
feature. Indeed, if our concern is to provide ethical standards for
assessing and designing recommendations for \emph{users}, then our
foundations should not exclude trading-off the good of the users for the
good of other stakeholders. We may have reasons to not design
recommendations systems that users can game, but these reasons are
external to actionability and user well-being.

\hypertarget{fairness}{%
\subsection{Fairness}\label{fairness}}

One important motivation for making sure that recommendations are
actionable is that some recommendations may be actionable for one person
and not for another. However, mere actionability may not capture all the
features we want from good recommendations. A recommendation may be
actionable for two different people yet differ in their cost. This raises issues of fairness, especially if the
grounds for the cost are unjust. Recommendations that are more costly
for particular groups or communities may signal that there is
discrimination. For example, just recommendations to acquire more work experience may ignore various work and care responsibilities that differ between groups. If we want recourse to be fair,
we thus need an account of recourse fairness.

\citet{Gupta:2019} propose to measure recourse fairness in terms of the
average group distance to the decision boundary. However, as
\citet{vonKugelgen:2021} note, distance-based notions do not take into
account the real causal effects---and thus costs---of intervening on
variables. Accordingly, they suggest an individual and group-level
\emph{causal} notion of recourse fairness. Although arguably a step in
the right direction, a causal approach faces several obstacles. One is
that thinking of discrimination in causal terms is contentious (see
sec.~5.1 above).\footnote{They also propose to improve recourse fairness
  through ``societal interventions''. However, these interventions
  are not easily available to individuals seeking recourse and it is thus
  unclear why they should qualify as recommendations in our sense.}
Another more serious issue is that causal reasoning will not tell, by
itself, what causes \emph{should} count. For instance, some theories of
justice consider that burdens and benefits should be distributed
according to desert \citep{Brouwer:2019}. A recommendation might
be costly for a person, but she might \emph{deserve} to be in that
position. One might argue that the proverbial surfer failing to save
should perhaps not obtain a loan so easily.

Although the capability approach does not solve by itself all issues
related to algorithmic fairness, it provides a theoretical framework
within which to conceptualize these problems. Someone more interested in
the fairness of outcomes could try to optimize for recommendations that
provide fair functionings; others more interested in opportunities may
instead consider that capabilities should be the key metric of justice.
And the notion of `conversion factors' provides a language to formulate
various issues related to fairness. Social conversion factors can be
social norms that discriminate and personal conversion factors such as
having a disability may justify compensating people seeking recourse.

\hypertarget{conclusion}{%
\section{Conclusion}\label{conclusion}}

Designers of algorithmic systems are often interested in providing
recourse to users, viz.~the ability to obtain or reverse a decision from
a model. Recourse has often been associated with providing
counterfactual explanations. We first proposed to reframe recourse
not as an explanation problem, but as a recommendation problem. The aim
of recourse is not necessarily to understand why the model made the
decision, but rather simply to allow users to achieve results
they value. Not all explanations provide recourse and not all
recommendations provide understanding. One benefit of viewing recourse
as a recommendation problem is that it leverages the existing
literature on recommendation systems. But it also creates a challenge
for designers of these systems: What are good recommendations?

We argued that the capability approach provides plausible and fruitful
ethical standards for the design of recommendation systems whose goal is
to give recourse to users. The capability approach is both descriptively
and normatively adequate; it captures the relevant features of recourse
and provides an ethical justification for why some recommendations are
better than others. In particular, we submitted that good
recommendations will be those that promote people's \emph{capabilities}.
To illustrate the relevance of the framework, we discussed one
particular constraint to recourse, diversity. We closed by discussing
several implications of adopting the capability approach for recourse
research beyond diversity.

To conclude, we would like to emphasize that the capability approach is not the only framework which can be used to conceptualize the
ethical constraints to recourse. Although there might be other suitable
alternatives in some contexts, we simply hold that the capability approach is a
worthy contender. That being said, one important message we hope our
discussion conveyed is that if recourse is to live up to its ethical
promise, then we cannot dispense with examining the ethical assumptions
underlying what we take good recommendations to be.

\begin{acks}
This work is supported by the Netherlands Organization for Scientific Research (NWO grant number VI.Veni.201F.051), and part of the research programme Ethics of Socially Disruptive Technologies, which is funded by the Gravitation programme of the Dutch Ministry of Education, Culture, and Science and the Netherlands Organization for Scientific Research (NWO grant number 024.004.031). The authors discussed this work with Maastricht University's xAI research group, the ESDiT Society Line, at TU Dortmund, ACFAS 2022, and the ECPAI ML Opacity Circle. We thank the participants for comments on previous versions of the manuscript.
\end{acks}

\bibliographystyle{ACM-Reference-Format}
\bibliography{AIES2022.bbl}


\begin{thebibliography}{62}


\ifx \showCODEN    \undefined \def \showCODEN     #1{\unskip}     \fi
\ifx \showDOI      \undefined \def \showDOI       #1{#1}\fi
\ifx \showISBNx    \undefined \def \showISBNx     #1{\unskip}     \fi
\ifx \showISBNxiii \undefined \def \showISBNxiii  #1{\unskip}     \fi
\ifx \showISSN     \undefined \def \showISSN      #1{\unskip}     \fi
\ifx \showLCCN     \undefined \def \showLCCN      #1{\unskip}     \fi
\ifx \shownote     \undefined \def \shownote      #1{#1}          \fi
\ifx \showarticletitle \undefined \def \showarticletitle #1{#1}   \fi
\ifx \showURL      \undefined \def \showURL       {\relax}        \fi
\providecommand\bibfield[2]{#2}
\providecommand\bibinfo[2]{#2}
\providecommand\natexlab[1]{#1}
\providecommand\showeprint[2][]{arXiv:#2}

\bibitem[{Al-Ajlani} et~al\mbox{.}(2020)]%
        {Al-Ajlani:2020}
\bibfield{author}{\bibinfo{person}{Haya {Al-Ajlani}}, \bibinfo{person}{Luc
  Van~Ootegem}, {and} \bibinfo{person}{Elsy Verhofstadt}.}
  \bibinfo{year}{2020}\natexlab{}.
\newblock \showarticletitle{Does {{Well-Being Vary}} with an
  {{Individual-Specific Weighting Scheme}}?}
\newblock \bibinfo{journal}{\emph{Applied Research in Quality of Life}}
  \bibinfo{volume}{15}, \bibinfo{number}{5} (\bibinfo{date}{Nov.}
  \bibinfo{year}{2020}), \bibinfo{pages}{1285--1302}.
\newblock
\showISSN{1871-2576}
\urldef\tempurl%
\url{https://doi.org/10.1007/s11482-019-09733-0}
\showDOI{\tempurl}


\bibitem[Alkire(2013)]%
        {Alkire:2013}
\bibfield{author}{\bibinfo{person}{Sabina Alkire}.}
  \bibinfo{year}{2013}\natexlab{}.
\newblock \showarticletitle{Choosing {{Dimensions}}: {{The Capability
  Approach}} and {{Multidimensional Poverty}}}.
\newblock In \bibinfo{booktitle}{\emph{The {{Many Dimensions}} of
  {{Poverty}}}}, \bibfield{editor}{\bibinfo{person}{Nanak Kakwani} {and}
  \bibinfo{person}{Jacques Silber}} (Eds.). \bibinfo{publisher}{{Palgrave
  Macmillan UK}}, \bibinfo{address}{{London}}, \bibinfo{pages}{89--119}.
\newblock
\showISBNx{978-0-230-59240-7}
\urldef\tempurl%
\url{https://doi.org/10.1057/9780230592407_6}
\showDOI{\tempurl}


\bibitem[Alkire et~al\mbox{.}(2015)]%
        {Alkire:2015}
\bibfield{author}{\bibinfo{person}{Sabina Alkire},
  \bibinfo{person}{Jos{\'e}~Manuel Roche}, \bibinfo{person}{Paola Ballon},
  \bibinfo{person}{James Foster}, \bibinfo{person}{Maria~Emma Santos}, {and}
  \bibinfo{person}{Suman Seth}.} \bibinfo{year}{2015}\natexlab{}.
\newblock \bibinfo{booktitle}{\emph{Multidimensional {{Poverty Measurement}}
  and {{Analysis}}}}.
\newblock \bibinfo{publisher}{{Oxford University Press}},
  \bibinfo{address}{{Oxford}}.
\newblock
\showISBNx{978-0-19-968949-1}


\bibitem[Angwin et~al\mbox{.}(2016)]%
        {Angwin:2016}
\bibfield{author}{\bibinfo{person}{Julia Angwin}, \bibinfo{person}{Jeff
  Larson}, \bibinfo{person}{Surya Mattu}, \bibinfo{person}{Lauren Kirchner},
  {and} \bibinfo{person}{Probublica}.} \bibinfo{year}{2016}\natexlab{}.
\newblock \bibinfo{title}{Machine {{Bias}}. {{There}}'s Software Used across
  the Country to Predict Future Criminals. {{And}} It's Biased against Blacks}.
\newblock
  \bibinfo{howpublished}{https://www.propublica.org/article/machine-bias-risk-assessments-in-criminal-sentencing}.
\newblock


\bibitem[Barocas et~al\mbox{.}(2020)]%
        {Barocas:2020a}
\bibfield{author}{\bibinfo{person}{Solon Barocas}, \bibinfo{person}{Andrew~D.
  Selbst}, {and} \bibinfo{person}{Manish Raghavan}.}
  \bibinfo{year}{2020}\natexlab{}.
\newblock \showarticletitle{The Hidden Assumptions behind Counterfactual
  Explanations and Principal Reasons}. In \bibinfo{booktitle}{\emph{Proceedings
  of the 2020 {{Conference}} on {{Fairness}}, {{Accountability}}, and
  {{Transparency}}}} \emph{(\bibinfo{series}{{{FAT}}* '20})}.
  \bibinfo{publisher}{{Association for Computing Machinery}},
  \bibinfo{address}{{New York, NY, USA}}, \bibinfo{pages}{80--89}.
\newblock
\showISBNx{978-1-4503-6936-7}
\urldef\tempurl%
\url{https://doi.org/10.1145/3351095.3372830}
\showDOI{\tempurl}


\bibitem[Bright et~al\mbox{.}(2015)]%
        {Bright:2015}
\bibfield{author}{\bibinfo{person}{Liam~Kofi Bright}, \bibinfo{person}{Daniel
  Malinsky}, {and} \bibinfo{person}{Morgan Thompson}.}
  \bibinfo{year}{2015}\natexlab{}.
\newblock \showarticletitle{Causally {{Interpreting Intersectionality
  Theory}}}.
\newblock \bibinfo{journal}{\emph{Philosophy of Science}} \bibinfo{volume}{83},
  \bibinfo{number}{1} (\bibinfo{date}{Dec.} \bibinfo{year}{2015}),
  \bibinfo{pages}{60--81}.
\newblock
\showISSN{0031-8248}
\urldef\tempurl%
\url{https://doi.org/10.1086/684173}
\showDOI{\tempurl}


\bibitem[Brouwer and Mulligan(2019)]%
        {Brouwer:2019}
\bibfield{author}{\bibinfo{person}{Huub Brouwer} {and} \bibinfo{person}{Thomas
  Mulligan}.} \bibinfo{year}{2019}\natexlab{}.
\newblock \showarticletitle{Why Not Be a Desertist?}
\newblock \bibinfo{journal}{\emph{Philosophical Studies}}
  \bibinfo{volume}{176}, \bibinfo{number}{9} (\bibinfo{date}{Sept.}
  \bibinfo{year}{2019}), \bibinfo{pages}{2271--2288}.
\newblock
\showISSN{1573-0883}
\urldef\tempurl%
\url{https://doi.org/10.1007/s11098-018-1125-4}
\showDOI{\tempurl}


\bibitem[Castagnos et~al\mbox{.}(2013)]%
        {Castagnos:2013}
\bibfield{author}{\bibinfo{person}{Sylvain Castagnos}, \bibinfo{person}{Armelle
  Brun}, {and} \bibinfo{person}{Anne Boyer}.} \bibinfo{year}{2013}\natexlab{}.
\newblock \showarticletitle{When {{Diversity Is Needed}}... {{But Not
  Expected}}!}. In \bibinfo{booktitle}{\emph{International {{Conference}} on
  {{Advances}} in {{Information Mining}} and {{Management}}}}.
  \bibinfo{publisher}{{IARIA XPS Press}}, \bibinfo{pages}{44}.
\newblock


\bibitem[Dandl et~al\mbox{.}(2020)]%
        {Dandl:2020}
\bibfield{author}{\bibinfo{person}{Susanne Dandl}, \bibinfo{person}{Christoph
  Molnar}, \bibinfo{person}{Martin Binder}, {and} \bibinfo{person}{Bernd
  Bischl}.} \bibinfo{year}{2020}\natexlab{}.
\newblock \showarticletitle{Multi-{{Objective Counterfactual Explanations}}}.
  In \bibinfo{booktitle}{\emph{Parallel {{Problem Solving}} from {{Nature}}
  \textendash{} {{PPSN XVI}}}} \emph{(\bibinfo{series}{Lecture {{Notes}} in
  {{Computer Science}}})}, \bibfield{editor}{\bibinfo{person}{Thomas B{\"a}ck},
  \bibinfo{person}{Mike Preuss}, \bibinfo{person}{Andr{\'e} Deutz},
  \bibinfo{person}{Hao Wang}, \bibinfo{person}{Carola Doerr},
  \bibinfo{person}{Michael Emmerich}, {and} \bibinfo{person}{Heike Trautmann}}
  (Eds.). \bibinfo{publisher}{{Springer International Publishing}},
  \bibinfo{address}{{Cham}}, \bibinfo{pages}{448--469}.
\newblock
\showISBNx{978-3-030-58112-1}
\urldef\tempurl%
\url{https://doi.org/10.1007/978-3-030-58112-1_31}
\showDOI{\tempurl}


\bibitem[Dean et~al\mbox{.}(2020)]%
        {Dean:2020}
\bibfield{author}{\bibinfo{person}{Sarah Dean}, \bibinfo{person}{Sarah Rich},
  {and} \bibinfo{person}{Benjamin Recht}.} \bibinfo{year}{2020}\natexlab{}.
\newblock \showarticletitle{Recommendations and User Agency: The Reachability
  of Collaboratively-Filtered Information}. In
  \bibinfo{booktitle}{\emph{Proceedings of the 2020 {{Conference}} on
  {{Fairness}}, {{Accountability}}, and {{Transparency}}}}
  \emph{(\bibinfo{series}{{{FAT}}* '20})}. \bibinfo{publisher}{{Association for
  Computing Machinery}}, \bibinfo{address}{{New York, NY, USA}},
  \bibinfo{pages}{436--445}.
\newblock
\showISBNx{978-1-4503-6936-7}
\urldef\tempurl%
\url{https://doi.org/10.1145/3351095.3372866}
\showDOI{\tempurl}


\bibitem[Dressel and Farid(2018)]%
        {Dressel:2018}
\bibfield{author}{\bibinfo{person}{Julia Dressel} {and} \bibinfo{person}{Hany
  Farid}.} \bibinfo{year}{2018}\natexlab{}.
\newblock \showarticletitle{The Accuracy, Fairness, and Limits of Predicting
  Recidivism}.
\newblock \bibinfo{journal}{\emph{Science Advances}} \bibinfo{volume}{4},
  \bibinfo{number}{1} (\bibinfo{date}{Jan.} \bibinfo{year}{2018}),
  \bibinfo{pages}{eaao5580}.
\newblock
\urldef\tempurl%
\url{https://doi.org/10.1126/sciadv.aao5580}
\showDOI{\tempurl}


\bibitem[Friedman(1974)]%
        {Friedman:1974fk}
\bibfield{author}{\bibinfo{person}{Michael Friedman}.}
  \bibinfo{year}{1974}\natexlab{}.
\newblock \showarticletitle{Explanation and {{Scientific Understanding}}}.
\newblock \bibinfo{journal}{\emph{Journal of Philosophy}} \bibinfo{volume}{71},
  \bibinfo{number}{1} (\bibinfo{year}{1974}), \bibinfo{pages}{5--19}.
\newblock


\bibitem[Gaertner and Xu(2006)]%
        {Gaertner:2006}
\bibfield{author}{\bibinfo{person}{Wulf Gaertner} {and}
  \bibinfo{person}{Yongsheng Xu}.} \bibinfo{year}{2006}\natexlab{}.
\newblock \showarticletitle{Capability {{Sets}} as the {{Basis}} of a {{New
  Measure}} of {{Human Development}}}.
\newblock \bibinfo{journal}{\emph{Journal of Human Development}}
  \bibinfo{volume}{7}, \bibinfo{number}{3} (\bibinfo{date}{Nov.}
  \bibinfo{year}{2006}), \bibinfo{pages}{311--321}.
\newblock
\showISSN{1464-9888}
\urldef\tempurl%
\url{https://doi.org/10.1080/14649880600815891}
\showDOI{\tempurl}


\bibitem[Glymour and Glymour(2014)]%
        {Glymour:2014}
\bibfield{author}{\bibinfo{person}{Clark Glymour} {and}
  \bibinfo{person}{Madelyn~R. Glymour}.} \bibinfo{year}{2014}\natexlab{}.
\newblock \showarticletitle{Commentary: {{Race}} and {{Sex Are Causes}}}.
\newblock \bibinfo{journal}{\emph{Epidemiology}} \bibinfo{volume}{25},
  \bibinfo{number}{4} (\bibinfo{year}{2014}), \bibinfo{pages}{488--490}.
\newblock
\showISSN{1044-3983}


\bibitem[Grath et~al\mbox{.}(2018)]%
        {Grath:2018}
\bibfield{author}{\bibinfo{person}{Rory~Mc Grath}, \bibinfo{person}{Luca
  Costabello}, \bibinfo{person}{Chan~Le Van}, \bibinfo{person}{Paul Sweeney},
  \bibinfo{person}{Farbod Kamiab}, \bibinfo{person}{Zhao Shen}, {and}
  \bibinfo{person}{Freddy Lecue}.} \bibinfo{year}{2018}\natexlab{}.
\newblock \showarticletitle{Interpretable {{Credit Application Predictions With
  Counterfactual Explanations}}}.
\newblock \bibinfo{journal}{\emph{arXiv:1811.05245 [cs]}} (\bibinfo{date}{Nov.}
  \bibinfo{year}{2018}).
\newblock
\showeprint[arxiv]{1811.05245}~[cs]


\bibitem[Grimm(2010)]%
        {Grimm:2010rt}
\bibfield{author}{\bibinfo{person}{Stephen~R. Grimm}.}
  \bibinfo{year}{2010}\natexlab{}.
\newblock \showarticletitle{The Goal of Explanation}.
\newblock \bibinfo{journal}{\emph{Studies In History and Philosophy of Science
  Part A}} \bibinfo{volume}{41}, \bibinfo{number}{4} (\bibinfo{year}{2010}),
  \bibinfo{pages}{337--344}.
\newblock
\showISSN{0039-3681}
\urldef\tempurl%
\url{https://doi.org/10.1016/j.shpsa.2010.10.006}
\showDOI{\tempurl}


\bibitem[Gupta et~al\mbox{.}(2019)]%
        {Gupta:2019}
\bibfield{author}{\bibinfo{person}{Vivek Gupta}, \bibinfo{person}{Pegah
  Nokhiz}, \bibinfo{person}{Chitradeep~Dutta Roy}, {and}
  \bibinfo{person}{Suresh Venkatasubramanian}.}
  \bibinfo{year}{2019}\natexlab{}.
\newblock \showarticletitle{Equalizing {{Recourse}} across {{Groups}}}.
\newblock \bibinfo{journal}{\emph{arXiv:1909.03166 [cs, stat]}}
  (\bibinfo{date}{Sept.} \bibinfo{year}{2019}).
\newblock
\showeprint[arxiv]{1909.03166}~[cs, stat]


\bibitem[Hanna et~al\mbox{.}(2020)]%
        {Hanna:2020}
\bibfield{author}{\bibinfo{person}{Alex Hanna}, \bibinfo{person}{Emily Denton},
  \bibinfo{person}{Andrew Smart}, {and} \bibinfo{person}{Jamila {Smith-Loud}}.}
  \bibinfo{year}{2020}\natexlab{}.
\newblock \showarticletitle{Towards a Critical Race Methodology in Algorithmic
  Fairness}. In \bibinfo{booktitle}{\emph{Proceedings of the 2020
  {{Conference}} on {{Fairness}}, {{Accountability}}, and {{Transparency}}}}
  \emph{(\bibinfo{series}{{{FAT}}* '20})}. \bibinfo{publisher}{{Association for
  Computing Machinery}}, \bibinfo{address}{{New York, NY, USA}},
  \bibinfo{pages}{501--512}.
\newblock
\showISBNx{978-1-4503-6936-7}
\urldef\tempurl%
\url{https://doi.org/10.1145/3351095.3372826}
\showDOI{\tempurl}


\bibitem[Hu and Pu(2011)]%
        {Hu:2011}
\bibfield{author}{\bibinfo{person}{Rong Hu} {and} \bibinfo{person}{Pearl Pu}.}
  \bibinfo{year}{2011}\natexlab{}.
\newblock \showarticletitle{Helping {{Users Perceive Recommendation
  Diversity}}}. In \bibinfo{booktitle}{\emph{Proceedings of the {{Workshop}} on
  {{Novelty}} and {{Diversity}} in {{Recommender Systems}} ({{DiveRS}} 2011)}}.
  \bibinfo{address}{{Chicago, Illinois, USA}}, \bibinfo{pages}{43--50}.
\newblock


\bibitem[Joshi et~al\mbox{.}(2019)]%
        {Joshi:2019}
\bibfield{author}{\bibinfo{person}{Shalmali Joshi}, \bibinfo{person}{Oluwasanmi
  Koyejo}, \bibinfo{person}{Warut Vijitbenjaronk}, \bibinfo{person}{Been Kim},
  {and} \bibinfo{person}{Joydeep Ghosh}.} \bibinfo{year}{2019}\natexlab{}.
\newblock \showarticletitle{Towards {{Realistic Individual Recourse}} and
  {{Actionable Explanations}} in {{Black-Box Decision Making Systems}}}.
\newblock \bibinfo{journal}{\emph{arXiv:1907.09615 [cs, stat]}}
  (\bibinfo{date}{July} \bibinfo{year}{2019}).
\newblock
\showeprint[arxiv]{1907.09615}~[cs, stat]


\bibitem[Karimi et~al\mbox{.}(2021a)]%
        {Karimi:2021a}
\bibfield{author}{\bibinfo{person}{Amir-Hossein Karimi},
  \bibinfo{person}{Gilles Barthe}, \bibinfo{person}{Bernhard Sch{\"o}lkopf},
  {and} \bibinfo{person}{Isabel Valera}.} \bibinfo{year}{2021}\natexlab{a}.
\newblock \showarticletitle{A Survey of Algorithmic Recourse: Definitions,
  Formulations, Solutions, and Prospects}.
\newblock \bibinfo{journal}{\emph{arXiv:2010.04050 [cs, stat]}}
  (\bibinfo{date}{March} \bibinfo{year}{2021}).
\newblock
\showeprint[arxiv]{2010.04050}~[cs, stat]


\bibitem[Karimi et~al\mbox{.}(2021b)]%
        {Karimi:2021}
\bibfield{author}{\bibinfo{person}{Amir-Hossein Karimi},
  \bibinfo{person}{Bernhard Sch{\"o}lkopf}, {and} \bibinfo{person}{Isabel
  Valera}.} \bibinfo{year}{2021}\natexlab{b}.
\newblock \showarticletitle{Algorithmic {{Recourse}}: From {{Counterfactual
  Explanations}} to {{Interventions}}}. In
  \bibinfo{booktitle}{\emph{Proceedings of the 2021 {{ACM Conference}} on
  {{Fairness}}, {{Accountability}}, and {{Transparency}}}}
  \emph{(\bibinfo{series}{{{FAccT}} '21})}. \bibinfo{publisher}{{Association
  for Computing Machinery}}, \bibinfo{address}{{New York, NY, USA}},
  \bibinfo{pages}{353--362}.
\newblock
\showISBNx{978-1-4503-8309-7}
\urldef\tempurl%
\url{https://doi.org/10.1145/3442188.3445899}
\showDOI{\tempurl}


\bibitem[Karimi et~al\mbox{.}(2020)]%
        {Karimi:2020}
\bibfield{author}{\bibinfo{person}{Amir-Hossein Karimi},
  \bibinfo{person}{Julius {von K{\"u}gelgen}}, \bibinfo{person}{Bernhard
  Sch{\"o}lkopf}, {and} \bibinfo{person}{Isabel Valera}.}
  \bibinfo{year}{2020}\natexlab{}.
\newblock \showarticletitle{Algorithmic Recourse under Imperfect Causal
  Knowledge: A Probabilistic Approach}.
\newblock \bibinfo{journal}{\emph{arXiv:2006.06831 [cs, stat]}}
  (\bibinfo{date}{Oct.} \bibinfo{year}{2020}).
\newblock
\showeprint[arxiv]{2006.06831}~[cs, stat]


\bibitem[Kasirzadeh and Smart(2021)]%
        {Kasirzadeh:2021}
\bibfield{author}{\bibinfo{person}{Atoosa Kasirzadeh} {and}
  \bibinfo{person}{Andrew Smart}.} \bibinfo{year}{2021}\natexlab{}.
\newblock \showarticletitle{The {{Use}} and {{Misuse}} of {{Counterfactuals}}
  in {{Ethical Machine Learning}}}. In \bibinfo{booktitle}{\emph{Proceedings of
  the 2021 {{ACM Conference}} on {{Fairness}}, {{Accountability}}, and
  {{Transparency}}}} \emph{(\bibinfo{series}{{{FAccT}} '21})}.
  \bibinfo{publisher}{{Association for Computing Machinery}},
  \bibinfo{address}{{New York, NY, USA}}, \bibinfo{pages}{228--236}.
\newblock
\showISBNx{978-1-4503-8309-7}
\urldef\tempurl%
\url{https://doi.org/10.1145/3442188.3445886}
\showDOI{\tempurl}


\bibitem[K{\"o}nig et~al\mbox{.}(2021)]%
        {Konig:2021}
\bibfield{author}{\bibinfo{person}{Gunnar K{\"o}nig}, \bibinfo{person}{Timo
  Freiesleben}, {and} \bibinfo{person}{Moritz {Grosse-Wentrup}}.}
  \bibinfo{year}{2021}\natexlab{}.
\newblock \showarticletitle{A {{Causal Perspective}} on {{Meaningful}} and
  {{Robust Algorithmic Recourse}}}.
\newblock \bibinfo{journal}{\emph{arXiv:2107.07853 [cs, stat]}}
  (\bibinfo{date}{July} \bibinfo{year}{2021}).
\newblock
\showeprint[arxiv]{2107.07853}~[cs, stat]


\bibitem[Kunaver and Po{\v z}rl(2017)]%
        {Kunaver:2017}
\bibfield{author}{\bibinfo{person}{Matev{\v z} Kunaver} {and}
  \bibinfo{person}{Toma{\v z} Po{\v z}rl}.} \bibinfo{year}{2017}\natexlab{}.
\newblock \showarticletitle{Diversity in Recommender Systems \textendash{}
  {{A}} Survey}.
\newblock \bibinfo{journal}{\emph{Knowledge-Based Systems}}
  \bibinfo{volume}{123} (\bibinfo{date}{May} \bibinfo{year}{2017}),
  \bibinfo{pages}{154--162}.
\newblock
\showISSN{0950-7051}
\urldef\tempurl%
\url{https://doi.org/10.1016/j.knosys.2017.02.009}
\showDOI{\tempurl}


\bibitem[Lakkaraju(2021)]%
        {Lakkaraju:2021a}
\bibfield{author}{\bibinfo{person}{Himabindu Lakkaraju}.}
  \bibinfo{year}{2021}\natexlab{}.
\newblock \showarticletitle{Towards {{Reliable}} and {{Practicable Algorithmic
  Recourse}}}. In \bibinfo{booktitle}{\emph{Proceedings of the 30th {{ACM
  International Conference}} on {{Information}} \& {{Knowledge Management}}}}.
  \bibinfo{publisher}{{Association for Computing Machinery}},
  \bibinfo{address}{{New York, NY, USA}}, \bibinfo{pages}{4}.
\newblock
\showISBNx{978-1-4503-8446-9}


\bibitem[Lipton(2018)]%
        {Lipton:2018}
\bibfield{author}{\bibinfo{person}{Zachary~C Lipton}.}
  \bibinfo{year}{2018}\natexlab{}.
\newblock \showarticletitle{The {{Mythos}} of {{Model Interpretability}}:
  {{In}} Machine Learning, the Concept of Interpretability Is Both Important
  and Slippery}.
\newblock \bibinfo{journal}{\emph{Queue}} \bibinfo{volume}{16},
  \bibinfo{number}{3} (\bibinfo{year}{2018}), \bibinfo{pages}{31--57}.
\newblock
\urldef\tempurl%
\url{https://doi.org/10.1145/3236386.3241340}
\showDOI{\tempurl}


\bibitem[Loecherbach et~al\mbox{.}(2020)]%
        {Loecherbach:2020}
\bibfield{author}{\bibinfo{person}{Felicia Loecherbach},
  \bibinfo{person}{Judith Moeller}, \bibinfo{person}{Damian Trilling}, {and}
  \bibinfo{person}{Wouter {van Atteveldt}}.} \bibinfo{year}{2020}\natexlab{}.
\newblock \showarticletitle{The {{Unified Framework}} of {{Media Diversity}}:
  {{A Systematic Literature Review}}}.
\newblock \bibinfo{journal}{\emph{Digital Journalism}} \bibinfo{volume}{8},
  \bibinfo{number}{5} (\bibinfo{date}{May} \bibinfo{year}{2020}),
  \bibinfo{pages}{605--642}.
\newblock
\showISSN{2167-0811}
\urldef\tempurl%
\url{https://doi.org/10.1080/21670811.2020.1764374}
\showDOI{\tempurl}


\bibitem[Marcellesi(2013)]%
        {Marcellesi:2013}
\bibfield{author}{\bibinfo{person}{Alexandre Marcellesi}.}
  \bibinfo{year}{2013}\natexlab{}.
\newblock \showarticletitle{Is {{Race}} a {{Cause}}?}
\newblock \bibinfo{journal}{\emph{Philosophy of Science}} \bibinfo{volume}{80},
  \bibinfo{number}{5} (\bibinfo{date}{Dec.} \bibinfo{year}{2013}),
  \bibinfo{pages}{650--659}.
\newblock
\showISSN{0031-8248, 1539-767X}
\urldef\tempurl%
\url{https://doi.org/10.1086/673721}
\showDOI{\tempurl}


\bibitem[Mothilal et~al\mbox{.}(2020)]%
        {Mothilal:2020}
\bibfield{author}{\bibinfo{person}{Ramaravind~K. Mothilal},
  \bibinfo{person}{Amit Sharma}, {and} \bibinfo{person}{Chenhao Tan}.}
  \bibinfo{year}{2020}\natexlab{}.
\newblock \showarticletitle{Explaining Machine Learning Classifiers through
  Diverse Counterfactual Explanations}. In
  \bibinfo{booktitle}{\emph{Proceedings of the 2020 {{Conference}} on
  {{Fairness}}, {{Accountability}}, and {{Transparency}}}}
  \emph{(\bibinfo{series}{{{FAT}}* '20})}. \bibinfo{publisher}{{Association for
  Computing Machinery}}, \bibinfo{address}{{New York, NY, USA}},
  \bibinfo{pages}{607--617}.
\newblock
\showISBNx{978-1-4503-6936-7}
\urldef\tempurl%
\url{https://doi.org/10.1145/3351095.3372850}
\showDOI{\tempurl}


\bibitem[Nussbaum(2000)]%
        {Nussbaum:2000}
\bibfield{author}{\bibinfo{person}{Martha~C. Nussbaum}.}
  \bibinfo{year}{2000}\natexlab{}.
\newblock \bibinfo{booktitle}{\emph{Women and {{Human Development}}: {{The
  Capabilities Approach}}}}.
\newblock \bibinfo{publisher}{{Cambridge University Press}},
  \bibinfo{address}{{Cambridge}}.
\newblock
\showISBNx{978-0-521-00385-8}


\bibitem[Oosterlaken(2009)]%
        {Oosterlaken:2009}
\bibfield{author}{\bibinfo{person}{Ilse Oosterlaken}.}
  \bibinfo{year}{2009}\natexlab{}.
\newblock \showarticletitle{Design for {{Development}}: {{A Capability
  Approach}}}.
\newblock \bibinfo{journal}{\emph{Design Issues}} \bibinfo{volume}{25},
  \bibinfo{number}{4} (\bibinfo{date}{Oct.} \bibinfo{year}{2009}),
  \bibinfo{pages}{91--102}.
\newblock
\showISSN{0747-9360}
\urldef\tempurl%
\url{https://doi.org/10.1162/desi.2009.25.4.91}
\showDOI{\tempurl}


\bibitem[Pettit(2015)]%
        {Pettit:2015}
\bibfield{author}{\bibinfo{person}{Philip Pettit}.}
  \bibinfo{year}{2015}\natexlab{}.
\newblock \bibinfo{booktitle}{\emph{The {{Robust Demands}} of the {{Good}}:
  {{Ethics}} with {{Attachment}}, {{Virtue}}, and {{Respect}}}}.
\newblock \bibinfo{publisher}{{Oxford University Press}}.
\newblock
\showISBNx{978-0-19-873260-0}


\bibitem[Rasouli and Yu(2021)]%
        {Rasouli:2021}
\bibfield{author}{\bibinfo{person}{Peyman Rasouli} {and}
  \bibinfo{person}{Ingrid~Chieh Yu}.} \bibinfo{year}{2021}\natexlab{}.
\newblock \showarticletitle{{{CARE}}: {{Coherent Actionable Recourse}} Based on
  {{Sound Counterfactual Explanations}}}.
\newblock \bibinfo{journal}{\emph{arXiv:2108.08197 [cs]}} (\bibinfo{date}{Aug.}
  \bibinfo{year}{2021}).
\newblock
\showeprint[arxiv]{2108.08197}~[cs]


\bibitem[Ribeiro et~al\mbox{.}(2016)]%
        {Ribeiro:2016}
\bibfield{author}{\bibinfo{person}{Marco~Tulio Ribeiro},
  \bibinfo{person}{Sameer Singh}, {and} \bibinfo{person}{Carlos Guestrin}.}
  \bibinfo{year}{2016}\natexlab{}.
\newblock \showarticletitle{"{{Why Should I Trust You}}?": {{Explaining}} the
  {{Predictions}} of {{Any Classifier}}}. In
  \bibinfo{booktitle}{\emph{Proceedings of the 22nd {{ACM SIGKDD International
  Conference}} on {{Knowledge Discovery}} and {{Data Mining}}}}
  \emph{(\bibinfo{series}{{{KDD}} '16})}. \bibinfo{publisher}{{Association for
  Computing Machinery}}, \bibinfo{address}{{New York, NY, USA}},
  \bibinfo{pages}{1135--1144}.
\newblock
\showISBNx{978-1-4503-4232-2}
\urldef\tempurl%
\url{https://doi.org/10.1145/2939672.2939778}
\showDOI{\tempurl}


\bibitem[Robeyns(2003)]%
        {Robeyns:2003}
\bibfield{author}{\bibinfo{person}{Ingrid Robeyns}.}
  \bibinfo{year}{2003}\natexlab{}.
\newblock \showarticletitle{Sen's {{Capability Approach}} and {{Gender
  Inequality}}: {{Selecting Relevant Capabilities}}}.
\newblock \bibinfo{journal}{\emph{Feminist Economics}} \bibinfo{volume}{9},
  \bibinfo{number}{2-3} (\bibinfo{date}{Jan.} \bibinfo{year}{2003}),
  \bibinfo{pages}{61--92}.
\newblock
\showISSN{1354-5701}
\urldef\tempurl%
\url{https://doi.org/10.1080/1354570022000078024}
\showDOI{\tempurl}


\bibitem[Robeyns(2017)]%
        {Robeyns:2017a}
\bibfield{author}{\bibinfo{person}{Ingrid Robeyns}.}
  \bibinfo{year}{2017}\natexlab{}.
\newblock \bibinfo{booktitle}{\emph{Wellbeing, {{Freedom}} and {{Social
  Justice}}: {{The Capability Approach Re-examined}}}}.
\newblock \bibinfo{publisher}{{Open Book Publishers}}.
\newblock
\showISBNx{978-1-78374-421-3}


\bibitem[Robeyns and Byskov(2021)]%
        {Robeyns:2021}
\bibfield{author}{\bibinfo{person}{Ingrid Robeyns} {and}
  \bibinfo{person}{Morten~Fibieger Byskov}.} \bibinfo{year}{2021}\natexlab{}.
\newblock \showarticletitle{The {{Capability Approach}}}.
\newblock In \bibinfo{booktitle}{\emph{The {{Stanford Encyclopedia}} of
  {{Philosophy}}} (\bibinfo{edition}{winter 2021} ed.)},
  \bibfield{editor}{\bibinfo{person}{Edward~N. Zalta}} (Ed.).
  \bibinfo{publisher}{{Metaphysics Research Lab, Stanford University}}.
\newblock


\bibitem[Russell(2019)]%
        {Russell:2019b}
\bibfield{author}{\bibinfo{person}{Chris Russell}.}
  \bibinfo{year}{2019}\natexlab{}.
\newblock \showarticletitle{Efficient {{Search}} for {{Diverse Coherent
  Explanations}}}. In \bibinfo{booktitle}{\emph{Proceedings of the
  {{Conference}} on {{Fairness}}, {{Accountability}}, and {{Transparency}}}}
  \emph{(\bibinfo{series}{{{FAT}}* '19})}. \bibinfo{publisher}{{Association for
  Computing Machinery}}, \bibinfo{address}{{New York, NY, USA}},
  \bibinfo{pages}{20--28}.
\newblock
\showISBNx{978-1-4503-6125-5}
\urldef\tempurl%
\url{https://doi.org/10.1145/3287560.3287569}
\showDOI{\tempurl}


\bibitem[Ryan et~al\mbox{.}(2015)]%
        {Ryan:2015}
\bibfield{author}{\bibinfo{person}{Jean Ryan}, \bibinfo{person}{Anders
  Wretstrand}, {and} \bibinfo{person}{Steven~M. Schmidt}.}
  \bibinfo{year}{2015}\natexlab{}.
\newblock \showarticletitle{Exploring Public Transport as an Element of Older
  Persons' Mobility: {{A Capability Approach}} Perspective}.
\newblock \bibinfo{journal}{\emph{Journal of Transport Geography}}
  \bibinfo{volume}{48} (\bibinfo{date}{Oct.} \bibinfo{year}{2015}),
  \bibinfo{pages}{105--114}.
\newblock
\showISSN{0966-6923}
\urldef\tempurl%
\url{https://doi.org/10.1016/j.jtrangeo.2015.08.016}
\showDOI{\tempurl}


\bibitem[Schein et~al\mbox{.}(2002)]%
        {Schein:2002}
\bibfield{author}{\bibinfo{person}{Andrew~I. Schein},
  \bibinfo{person}{Alexandrin Popescul}, \bibinfo{person}{Lyle~H. Ungar}, {and}
  \bibinfo{person}{David~M. Pennock}.} \bibinfo{year}{2002}\natexlab{}.
\newblock \showarticletitle{Methods and Metrics for Cold-Start
  Recommendations}. In \bibinfo{booktitle}{\emph{Proceedings of the 25th Annual
  International {{ACM SIGIR}} Conference on {{Research}} and Development in
  Information Retrieval}} \emph{(\bibinfo{series}{{{SIGIR}} '02})}.
  \bibinfo{publisher}{{Association for Computing Machinery}},
  \bibinfo{address}{{New York, NY, USA}}, \bibinfo{pages}{253--260}.
\newblock
\showISBNx{978-1-58113-561-9}
\urldef\tempurl%
\url{https://doi.org/10.1145/564376.564421}
\showDOI{\tempurl}


\bibitem[Sen(1979)]%
        {Sen:1979}
\bibfield{author}{\bibinfo{person}{Amartya Sen}.}
  \bibinfo{year}{1979}\natexlab{}.
\newblock \showarticletitle{Issues in the {{Measurement}} of {{Poverty}}}.
\newblock \bibinfo{journal}{\emph{The Scandinavian Journal of Economics}}
  \bibinfo{volume}{81}, \bibinfo{number}{2} (\bibinfo{year}{1979}),
  \bibinfo{pages}{285--307}.
\newblock
\showISSN{0347-0520}
\urldef\tempurl%
\url{https://doi.org/10.2307/3439966}
\showDOI{\tempurl}


\bibitem[Sen(1980)]%
        {Sen:1980fk}
\bibfield{author}{\bibinfo{person}{Amartya Sen}.}
  \bibinfo{year}{1980}\natexlab{}.
\newblock \showarticletitle{Equality of {{What}}?}
\newblock In \bibinfo{booktitle}{\emph{The {{Tanner Lecture}} on {{Human
  Values}}}}. Vol.~\bibinfo{volume}{1}. \bibinfo{publisher}{{Cambridge
  University Press}}, \bibinfo{address}{{Cambridge}},
  \bibinfo{pages}{197--220}.
\newblock


\bibitem[Sen(1992)]%
        {Sen:1992}
\bibfield{author}{\bibinfo{person}{Amartya Sen}.}
  \bibinfo{year}{1992}\natexlab{}.
\newblock \bibinfo{booktitle}{\emph{Inequality {{Reexamined}}}}.
\newblock \bibinfo{publisher}{{Harvard University Press}},
  \bibinfo{address}{{Cambridge, MA}}.
\newblock
\showISBNx{978-0-674-45255-8}


\bibitem[Sen(2009)]%
        {Sen:2009fk}
\bibfield{author}{\bibinfo{person}{Amartya Sen}.}
  \bibinfo{year}{2009}\natexlab{}.
\newblock \bibinfo{booktitle}{\emph{The {{Idea}} of {{Justice}}}}.
\newblock \bibinfo{publisher}{{The Belknap Press of Harvard University Press}},
  \bibinfo{address}{{Cambridge, MA}}.
\newblock


\bibitem[Simon et~al\mbox{.}(2013)]%
        {Simon:2013}
\bibfield{author}{\bibinfo{person}{Judit Simon}, \bibinfo{person}{Paul Anand},
  \bibinfo{person}{Alastair Gray}, \bibinfo{person}{Jorun Rugk{\aa}sa},
  \bibinfo{person}{Ksenija Yeeles}, {and} \bibinfo{person}{Tom Burns}.}
  \bibinfo{year}{2013}\natexlab{}.
\newblock \showarticletitle{Operationalising the Capability Approach for
  Outcome Measurement in Mental Health Research}.
\newblock \bibinfo{journal}{\emph{Social Science \& Medicine}}
  \bibinfo{volume}{98} (\bibinfo{date}{Dec.} \bibinfo{year}{2013}),
  \bibinfo{pages}{187--196}.
\newblock
\showISSN{0277-9536}
\urldef\tempurl%
\url{https://doi.org/10.1016/j.socscimed.2013.09.019}
\showDOI{\tempurl}


\bibitem[Stepin et~al\mbox{.}(2021)]%
        {Stepin:2021}
\bibfield{author}{\bibinfo{person}{Ilia Stepin}, \bibinfo{person}{Jose~M.
  Alonso}, \bibinfo{person}{Alejandro Catala}, {and}
  \bibinfo{person}{Mart{\'i}n {Pereira-Fari{\~n}a}}.}
  \bibinfo{year}{2021}\natexlab{}.
\newblock \showarticletitle{A {{Survey}} of {{Contrastive}} and
  {{Counterfactual Explanation Generation Methods}} for {{Explainable
  Artificial Intelligence}}}.
\newblock \bibinfo{journal}{\emph{IEEE Access}}  \bibinfo{volume}{9}
  (\bibinfo{year}{2021}), \bibinfo{pages}{11974--12001}.
\newblock
\showISSN{2169-3536}
\urldef\tempurl%
\url{https://doi.org/10.1109/ACCESS.2021.3051315}
\showDOI{\tempurl}


\bibitem[Sullivan et~al\mbox{.}(2019)]%
        {Sullivan:2019b}
\bibfield{author}{\bibinfo{person}{Emily Sullivan}, \bibinfo{person}{Dimitrios
  Bountouridis}, \bibinfo{person}{Jaron Harambam}, \bibinfo{person}{Shabnam
  Najafian}, \bibinfo{person}{Felicia Loecherbach}, \bibinfo{person}{Mykola
  Makhortykh}, \bibinfo{person}{Domokos Kelen}, \bibinfo{person}{Daricia
  Wilkinson}, \bibinfo{person}{David Graus}, {and} \bibinfo{person}{Nava
  Tintarev}.} \bibinfo{year}{2019}\natexlab{}.
\newblock \showarticletitle{Reading {{News}} with a {{Purpose}}: {{Explaining
  User Profiles}} for {{Self-Actualization}}}. In
  \bibinfo{booktitle}{\emph{Adjunct {{Publication}} of the 27th {{Conference}}
  on {{User Modeling}}, {{Adaptation}} and {{Personalization}}}}.
  \bibinfo{publisher}{{ACM}}, \bibinfo{address}{{Larnaca Cyprus}},
  \bibinfo{pages}{241--245}.
\newblock
\showISBNx{978-1-4503-6711-0}
\urldef\tempurl%
\url{https://doi.org/10.1145/3314183.3323456}
\showDOI{\tempurl}


\bibitem[Tintarev et~al\mbox{.}(2013)]%
        {Tintarev:2013}
\bibfield{author}{\bibinfo{person}{Nava Tintarev}, \bibinfo{person}{Matt
  Dennis}, {and} \bibinfo{person}{Judith Masthoff}.}
  \bibinfo{year}{2013}\natexlab{}.
\newblock \showarticletitle{Adapting {{Recommendation Diversity}} to
  {{Openness}} to {{Experience}}: {{A Study}} of {{Human Behaviour}}}. In
  \bibinfo{booktitle}{\emph{User {{Modeling}}, {{Adaptation}}, and
  {{Personalization}}}} \emph{(\bibinfo{series}{Lecture {{Notes}} in {{Computer
  Science}}})}, \bibfield{editor}{\bibinfo{person}{Sandra Carberry},
  \bibinfo{person}{Stephan Weibelzahl}, \bibinfo{person}{Alessandro Micarelli},
  {and} \bibinfo{person}{Giovanni Semeraro}} (Eds.).
  \bibinfo{publisher}{{Springer}}, \bibinfo{address}{{Berlin, Heidelberg}},
  \bibinfo{pages}{190--202}.
\newblock
\showISBNx{978-3-642-38844-6}
\urldef\tempurl%
\url{https://doi.org/10.1007/978-3-642-38844-6_16}
\showDOI{\tempurl}


\bibitem[Tintarev and Masthoff(2007)]%
        {Tintarev:2007}
\bibfield{author}{\bibinfo{person}{Nava Tintarev} {and} \bibinfo{person}{Judith
  Masthoff}.} \bibinfo{year}{2007}\natexlab{}.
\newblock \showarticletitle{A {{Survey}} of {{Explanations}} in {{Recommender
  Systems}}}. In \bibinfo{booktitle}{\emph{2007 {{IEEE}} 23rd {{International
  Conference}} on {{Data Engineering Workshop}}}}. \bibinfo{pages}{801--810}.
\newblock
\urldef\tempurl%
\url{https://doi.org/10.1109/ICDEW.2007.4401070}
\showDOI{\tempurl}


\bibitem[Ustun et~al\mbox{.}(2019)]%
        {Ustun:2019}
\bibfield{author}{\bibinfo{person}{Berk Ustun}, \bibinfo{person}{Alexander
  Spangher}, {and} \bibinfo{person}{Yang Liu}.}
  \bibinfo{year}{2019}\natexlab{}.
\newblock \showarticletitle{Actionable {{Recourse}} in {{Linear
  Classification}}}. In \bibinfo{booktitle}{\emph{Proceedings of the
  {{Conference}} on {{Fairness}}, {{Accountability}}, and {{Transparency}}}}
  \emph{(\bibinfo{series}{{{FAT}}* '19})}. \bibinfo{publisher}{{Association for
  Computing Machinery}}, \bibinfo{address}{{New York, NY, USA}},
  \bibinfo{pages}{10--19}.
\newblock
\showISBNx{978-1-4503-6125-5}
\urldef\tempurl%
\url{https://doi.org/10.1145/3287560.3287566}
\showDOI{\tempurl}


\bibitem[{van der Deijl}(2020)]%
        {vanderDeijl:2020a}
\bibfield{author}{\bibinfo{person}{Willem J.~A. {van der Deijl}}.}
  \bibinfo{year}{2020}\natexlab{}.
\newblock \showarticletitle{A {{Challenge}} for {{Capability Measures}} of
  {{Wellbeing}}}.
\newblock \bibinfo{journal}{\emph{Social Theory and Practice}}
  \bibinfo{volume}{46}, \bibinfo{number}{3} (\bibinfo{year}{2020}),
  \bibinfo{pages}{605--631}.
\newblock
\urldef\tempurl%
\url{https://doi.org/10.5840/soctheorpract202071799}
\showDOI{\tempurl}


\bibitem[Van~Ootegem and Verhofstadt(2012)]%
        {VanOotegem:2012}
\bibfield{author}{\bibinfo{person}{Luc Van~Ootegem} {and} \bibinfo{person}{Elsy
  Verhofstadt}.} \bibinfo{year}{2012}\natexlab{}.
\newblock \showarticletitle{Using {{Capabilities}} as an {{Alternative
  Indicator}} for {{Well-being}}}.
\newblock \bibinfo{journal}{\emph{Social Indicators Research}}
  \bibinfo{volume}{106}, \bibinfo{number}{1} (\bibinfo{date}{March}
  \bibinfo{year}{2012}), \bibinfo{pages}{133--152}.
\newblock
\showISSN{1573-0921}
\urldef\tempurl%
\url{https://doi.org/10.1007/s11205-011-9799-4}
\showDOI{\tempurl}


\bibitem[Venkatasubramanian and Alfano(2020)]%
        {Venkatasubramanian:2020}
\bibfield{author}{\bibinfo{person}{Suresh Venkatasubramanian} {and}
  \bibinfo{person}{Mark Alfano}.} \bibinfo{year}{2020}\natexlab{}.
\newblock \showarticletitle{The Philosophical Basis of Algorithmic Recourse}.
  In \bibinfo{booktitle}{\emph{Proceedings of the 2020 {{Conference}} on
  {{Fairness}}, {{Accountability}}, and {{Transparency}}}}
  \emph{(\bibinfo{series}{{{FAT}}* '20})}. \bibinfo{publisher}{{Association for
  Computing Machinery}}, \bibinfo{address}{{New York, NY, USA}},
  \bibinfo{pages}{284--293}.
\newblock
\showISBNx{978-1-4503-6936-7}
\urldef\tempurl%
\url{https://doi.org/10.1145/3351095.3372876}
\showDOI{\tempurl}


\bibitem[{von K{\"u}gelgen} et~al\mbox{.}(2021)]%
        {vonKugelgen:2021}
\bibfield{author}{\bibinfo{person}{Julius {von K{\"u}gelgen}},
  \bibinfo{person}{Amir-Hossein Karimi}, \bibinfo{person}{Umang Bhatt},
  \bibinfo{person}{Isabel Valera}, \bibinfo{person}{Adrian Weller}, {and}
  \bibinfo{person}{Bernhard Sch{\"o}lkopf}.} \bibinfo{year}{2021}\natexlab{}.
\newblock \showarticletitle{On the {{Fairness}} of {{Causal Algorithmic
  Recourse}}}.
\newblock \bibinfo{journal}{\emph{arXiv:2010.06529 [cs, stat]}}
  (\bibinfo{date}{June} \bibinfo{year}{2021}).
\newblock
\showeprint[arxiv]{2010.06529}~[cs, stat]


\bibitem[Vrijenhoek et~al\mbox{.}(2021)]%
        {Vrijenhoek:2021}
\bibfield{author}{\bibinfo{person}{Sanne Vrijenhoek}, \bibinfo{person}{Mesut
  Kaya}, \bibinfo{person}{Nadia Metoui}, \bibinfo{person}{Judith M{\"o}ller},
  \bibinfo{person}{Daan Odijk}, {and} \bibinfo{person}{Natali Helberger}.}
  \bibinfo{year}{2021}\natexlab{}.
\newblock \showarticletitle{Recommenders with a {{Mission}}: {{Assessing
  Diversity}} in {{News Recommendations}}}. In
  \bibinfo{booktitle}{\emph{Proceedings of the 2021 {{Conference}} on {{Human
  Information Interaction}} and {{Retrieval}}}}.
  \bibinfo{publisher}{{Association for Computing Machinery}},
  \bibinfo{address}{{New York, NY, USA}}, \bibinfo{pages}{173--183}.
\newblock
\showISBNx{978-1-4503-8055-3}


\bibitem[Wachter et~al\mbox{.}(2018)]%
        {Wachter:2018}
\bibfield{author}{\bibinfo{person}{Sandra Wachter}, \bibinfo{person}{Brent
  Mittelstadt}, {and} \bibinfo{person}{Chris Russell}.}
  \bibinfo{year}{2018}\natexlab{}.
\newblock \showarticletitle{Counterfactual {{Explanations}} without {{Opening}}
  the {{Black Box}}: {{Automated Decisions}} and the {{GDPR}}}.
\newblock \bibinfo{journal}{\emph{Harvard Journal of Law \& Technology (Harvard
  JOLT)}} \bibinfo{volume}{31}, \bibinfo{number}{2} (\bibinfo{year}{2018}),
  \bibinfo{pages}{841--887}.
\newblock


\bibitem[Walker and Unterhalter(2007)]%
        {Walker:2007}
\bibfield{author}{\bibinfo{person}{Melanie Walker} {and}
  \bibinfo{person}{Elaine Unterhalter}.} \bibinfo{year}{2007}\natexlab{}.
\newblock \showarticletitle{The {{Capability Approach}}: {{Its Potential}} for
  {{Work}} in {{Education}}}.
\newblock In \bibinfo{booktitle}{\emph{Amartya {{Sen}}'s {{Capability
  Approach}} and {{Social Justice}} in {{Education}}}},
  \bibfield{editor}{\bibinfo{person}{Melanie Walker} {and}
  \bibinfo{person}{Elaine Unterhalter}} (Eds.). \bibinfo{publisher}{{Palgrave
  Macmillan US}}, \bibinfo{address}{{New York}}, \bibinfo{pages}{1--18}.
\newblock
\showISBNx{978-0-230-60481-0}
\urldef\tempurl%
\url{https://doi.org/10.1057/9780230604810_1}
\showDOI{\tempurl}


\bibitem[Weinberger(2021)]%
        {Weinberger:2021b}
\bibfield{author}{\bibinfo{person}{Naftali Weinberger}.}
  \bibinfo{year}{2021}\natexlab{}.
\newblock \bibinfo{title}{Signal {{Manipulation}} and the {{Causal Status}} of
  {{Race}}}.
\newblock
\newblock
\urldef\tempurl%
\url{https://doi.org/10/1/Signal_Manipulation_and_the_Causal_Status_of_Race-7.pdf}
\showDOI{\tempurl}


\bibitem[Zednik(2021)]%
        {Zednik:2021}
\bibfield{author}{\bibinfo{person}{Carlos Zednik}.}
  \bibinfo{year}{2021}\natexlab{}.
\newblock \showarticletitle{Solving the {{Black Box Problem}}: {{A Normative
  Framework}} for {{Explainable Artificial Intelligence}}}.
\newblock \bibinfo{journal}{\emph{Philosophy \& Technology}}
  \bibinfo{volume}{34}, \bibinfo{number}{2} (\bibinfo{date}{June}
  \bibinfo{year}{2021}), \bibinfo{pages}{265--288}.
\newblock
\showISSN{2210-5441}
\urldef\tempurl%
\url{https://doi.org/10.1007/s13347-019-00382-7}
\showDOI{\tempurl}


\bibitem[Zheng(2009)]%
        {Zheng:2009}
\bibfield{author}{\bibinfo{person}{Yingqin Zheng}.}
  \bibinfo{year}{2009}\natexlab{}.
\newblock \showarticletitle{Different Spaces for E-Development: {{What}} Can We
  Learn from the Capability Approach?}
\newblock \bibinfo{journal}{\emph{Information Technology for Development}}
  \bibinfo{volume}{15}, \bibinfo{number}{2} (\bibinfo{year}{2009}),
  \bibinfo{pages}{66--82}.
\newblock
\showISSN{1554-0170}
\urldef\tempurl%
\url{https://doi.org/10.1002/itdj.20115}
\showDOI{\tempurl}


\end{thebibliography}

\end{document}